\documentclass[aps,prb,twocolumn,floatfix,amsmath,amssymb,
superscriptaddress]{revtex4-2}

\usepackage{graphicx}
\usepackage{bm}
\usepackage{booktabs}
\usepackage{array}
\usepackage{xcolor}
\usepackage[colorlinks,bookmarks=false,citecolor=blue,
linkcolor=red,urlcolor=blue]{hyperref}

\newcommand{\ket}[1]{\left|#1\right\rangle}
\newcommand{\bra}[1]{\left\langle#1\right|}
\newcommand{\braket}[2]{\left\langle#1\middle|#2\right\rangle}

\newcommand{\ii}{\mathrm{i}}

\newcommand{\id}{\mathbb{I}}

\newcommand{\plotfile}[2][]{%
  \IfFileExists{#2}{\includegraphics[#1]{#2}}{%
  \fbox{\parbox[c][0.14\textheight][c]{0.87\columnwidth}{%
    \centering\small Figure file not found:\\
 \texttt{\detokenize{#2}}}}}}

\begin{document}

\title{Boundaryless non-Hermitian skin effect and exceptional points in the Baxter--Fendley $Z_N$ spin chain}

\author{Xu Feng}
\affiliation{School of Physics, Peking University, Beijing 100871, China}
\email{xufeng98@pku.edu.cn}

\author{Shu Chen}
\affiliation{Beijing National Laboratory for Condensed Matter Physics,
Institute of Physics, Chinese Academy of Sciences, Beijing 100190, China}
\affiliation{School of Physical Sciences, University of Chinese Academy of
Sciences, Beijing 100049, China}
\email{schen@iphy.ac.cn}

\date{\today}

\begin{abstract}
The integrable non-Hermitian Baxter--Fendley $Z_N$ spin chain exhibits striking boundary sensitivity, where closing a single boundary bond alters bulk properties including the ground-state energy density and critical exponents. Here we uncover the physical mechanism underlying this sensitivity via a \emph{boundaryless} non-Hermitian skin effect (NHSE) in many-body configuration space and its interplay with integrability. Within each $Z_N$ charge sector, an $L$-site chain maps exactly onto a directed hopping model with a nonuniform complex potential on an $(L-1)$-dimensional periodic torus of linear size $N$. Crucially, this lattice has no geometric edge under either open (OBC) or periodic (PBC) boundary conditions; closing the chain simply adds a directed hopping channel at every site of the same torus. The interplay of directed hopping and the nonuniform potential drives spatial growth and decay of amplitudes, separating left and right eigenstates on this closed manifold under both OBC and PBC. Leveraging commuting conserved charges, we analytically resolve multichannel path interference on low-dimensional tori and track ground-state nonorthogonality in long chains well beyond exact diagonalization. Remarkably, whereas the OBC ground state exhibits rapid nonorthogonality growth ($\ln\kappa_0\propto L^2$) and an exponentially amplified closing-bond response, translation symmetry and self-duality bound the PBC single-bond response to $O(1)$ despite exponentially growing nonorthogonality ($\ln\kappa_0\propto L$). Continuously tuning the real boundary bond further drives excited states through exceptional points. Our results establish a unified configuration-space framework linking boundaryless NHSE, commuting conserved charges, and anomalous boundary sensitivity in non-Hermitian many-body systems.
\end{abstract}

\maketitle

\section{Introduction}
\label{sec:intro}

Non-Hermitian Hamiltonians arise naturally in effective descriptions of open quantum systems and exhibit novel phenomena that have no Hermitian counterparts~\cite{Feshbach1958,Feshbach1962,Rotter2009,ashida2020non,GongNH,KawabataNH,SatoNHSEReview,Znidaric1,Znidaric2,ChiralDamping,LSEYangFan,MISE,MISELongrange,MISEDPT,MISEDelocalization,hagaLiouvillianSkinEffect2021,LonghiSelfHealing,XueWenTanNonBlochEdgeDynamics,YangRealTimeEdgeDynamics,TrefethenEmbree,BosonLindbladian,xiaoanomalousentanglementscalingeigenvector,Li2026disorderinduced}. Two representative examples are exceptional points (EPs)~\cite{EPs,EPSensing,EPOptics2,PetermannFactorsEP,LiuChunHuiEP}, where eigenvalues and eigenvectors coalesce, and the non-Hermitian skin effect (NHSE), characterized by macroscopic eigenstate accumulation and strong sensitivity to boundary conditions~\cite{GBZ,BiorthogonalBBC,boundarysensitiveGuo,non-bloch,AnatomySkinModes,auxiliaryGBZ,PhaseSpaceGBZ,WindingSkin,WindingSkin2,SvdNHSE,CNHSE,GuoScaleFreeNHSE,zhangUniversalNonHermitianSkin2022,Amoeba,HuHighDimension,XingHu2026,MultifractalityNHSE,InteractingHN,ZhengMingChenNHSE,ManyBodyNHSEgaugecoupling,hamanakaInteractioninducedLiouvillianSkin,maoliouvillian2024,BoundarySensitiveLindb,LarsonLSE,HuYuMinManyBodyNHSE}. A simple illustration is the Hatano--Nelson model~\cite{HNmodel}: its asymmetric hopping gives extended Bloch states under periodic boundary conditions (PBC), whereas an extensive number of eigenstates localize near one end under open boundary conditions (OBC), accompanied by a dramatic change in the energy spectrum.

On another front, unusual boundary sensitivity has been found in the integrable Baxter--Fendley $Z_N$ spin chain~\cite{Baxter1989}. As a non-Hermitian generalization of the transverse-field Ising chain, this model is known for its free-parafermion spectrum and commuting conserved quantities under OBC~\cite{Fendley2014,AlcarazBatchelorLiu2017,BatchelorFreeParafermionReview,GraphFreeParafermion,Liu_2019,Pandey_DirtyFreeParafermion}. Notably, changing the physical boundary conditions from OBC to PBC alters not only the finite-size spectrum, but also bulk properties such as the ground-state energy density and critical exponents~\cite{AlcarazBatchelor2018,LiLiuBatchelor2026}. Although the model is exactly solvable, an intuitive physical explanation of this anomalous boundary sensitivity remains lacking.

This boundary sensitivity naturally suggests a connection to the NHSE. However, the standard single-particle picture---in which wavefunctions localize near an open physical boundary---does not directly apply to a many-body spin chain. To clarify the underlying mechanism, we map the Baxter--Fendley $Z_N$ chain exactly onto a directed hopping model in many-body configuration space. Within each fixed $Z_N$ charge sector, the effective lattice has $L-1$ independent directions, each containing $N$ sites. Thus, the chain length $L$ controls the dimension of configuration space, while the clock dimension $N$ sets its linear size along each direction. In this representation, the spin interactions act as directed hopping, while the transverse field produces a nonuniform, generally complex on-site potential.

This mapping reveals a \emph{boundaryless} NHSE. Here, the word ``boundaryless'' carries two distinct meanings. First, the configuration-space lattice is a periodic $(L-1)$-dimensional torus with no geometric boundary under either physical OBC or PBC: closing the spin chain does not create or remove any edge, but simply adds an extra directed hopping process at every site of the same torus. Second, anomalous eigenstate accumulation and exponentially small left--right overlaps occur under both physical OBC and PBC. Physically, non-Hermiticity ($H\neq H^\dagger$) alone is not sufficient for the NHSE; what is required is non-normality ($[H,H^\dagger]\neq 0$), which breaks eigenvector orthogonality. On a periodic torus, directed hopping alone is a normal operator ($[T,T^\dagger]=0$) and does not cause skin accumulation, whereas adding the nonuniform on-site potential breaks normality. Their combination produces spatially varying amplification and attenuation, allowing many eigenstates to accumulate in preferred subregions of a closed lattice.

A second focus of this work is the interplay between integrable conserved quantities and the NHSE across different scales. Rather than serving only as formal solutions, the commuting charges directly govern and constrain the NHSE. For $L=3$, where two incoming hopping paths interfere at each vertex of the two-dimensional torus, the OBC conserved charge $J_2$ locks their relative phase and predicts the skin-mode packet center and condition-number scaling analytically. For longer chains (large $L$ at fixed $N$), where high-dimensional wavefunctions are hard to visualize and localization depends on the basis, we quantify eigenvector nonorthogonality via the basis-invariant eigenvalue condition number $\kappa_n$~\cite{TrefethenEmbree,NonnormalHamOkuma,pseudospectraNHSE1,pseudospectraNHSE2,GeneralCriterionNHSE,FengNumericalInstability,scalingpseudospectraexponentiallysensitive,naves2026levelrepulsionfailsnonnormality}. Under OBC, the generating operator $\mathcal C(t)$ of the commuting charges yields an exact spectral projector $\Pi_0$ onto the ground state whose norm equals $\kappa_0^{\mathrm{OBC}}$. Because each factor $\mathcal C(t)$ admits an exact bond-dimension-two matrix product operator (MPO) representation, this enables matrix product state (MPS) calculations of $\kappa_0^{\mathrm{OBC}}$ well beyond exact diagonalization (ED). Moreover, the projector construction yields rigorous analytical upper bounds on OBC nonorthogonality---bounding all simple levels by $\ln\kappa_{\max}^{\mathrm{OBC}}=O(N)$ at fixed $L$ and the ground state by $\ln\kappa_0^{\mathrm{OBC}}=O(L^2)$ at fixed $N$. Complementarily, the exact OBC free-parafermion solution shows that the linear response $\kappa_{\mathrm b,0}^{\mathrm{OBC}}$ to the single closing bond grows exponentially with $L$, which directly connects the NHSE to anomalous boundary sensitivity and proves a rigorous exponential lower bound on $\kappa_0^{\mathrm{OBC}}$. Under PBC, although closing the chain breaks the free-parafermion structure of OBC, the periodic commuting-charge generating operator $\mathcal C_{\mathrm{PBC}}(t)$ decomposes into two bond-dimension-two open-chain MPO branches, providing a Trotter-error-free polynomial MPS filter for large $L$. Meanwhile, translation symmetry combined with self-duality strictly bounds the closing-bond response to $\kappa_{\mathrm b,0}^{\mathrm{PBC}}\le 1$ at the self-dual point even as $\kappa_0^{\mathrm{PBC}}$ grows exponentially with $L$. Finally, varying the real boundary coupling from OBC to PBC drives eigenstates to collide at EPs on the same periodic torus, where the same distinction between intrinsic nonorthogonality and physical boundary response becomes especially sharp.

The rest of the paper is organized as follows. In Sec.~\ref{sec:model}, we introduce the Baxter--Fendley $Z_N$ model, derive its exact mapping to the boundaryless configuration-space torus, and define the eigenvalue condition number, physical boundary response, and configuration-inversion symmetry. In Sec.~\ref{sec:skin}, we study the $L=2$ directed ring and the $L=3$ directed torus to illustrate the physical picture of boundaryless NHSE, basis duality under PBC, and two-channel interference fixed by the OBC conserved charge $J_2$. Section~\ref{sec:manybody} investigates longer chains at fixed $N$, presenting the commuting-charge MPO projector and polynomial filter methods, analytical bounds, and the contrast between OBC and PBC closing-bond responses. In Sec.~\ref{sec:ep}, we analyze the EPs induced by tuning the real boundary coupling, highlighting the contrast between generic intermediate EPs and duality-protected PBC endpoint EPs. Finally, Sec.~\ref{sec:conclusion} concludes the paper with a summary and discussion of open questions.

\section{Model and boundaryless configuration space}
\label{sec:model}

\subsection{Model and symmetries}
\label{subsec:model_symmetries}

We consider the Baxter--Fendley $Z_N$ spin chain with a tunable real boundary coupling $\mu$,
\begin{equation}
H(\lambda,\mu)
=-\sum_{j=1}^{L}X_j
-\lambda\sum_{j=1}^{L-1}Z_j^\dagger Z_{j+1}
-\mu Z_L^\dagger Z_1,
\label{eq:H_clean}
\end{equation}
where $L$ is the chain length, $N$ is the local clock dimension, and $\lambda>0$ is the bulk coupling. Here $\mu=0$ corresponds to OBC, and $\mu=\lambda$ corresponds to homogeneous PBC. The local clock operators act on the basis $\{\ket{m}\}_{m=0}^{N-1}$ (with labels taken modulo $N$) as
\begin{align}
Z\ket{m}&=\omega^m\ket{m}, & X\ket{m}&=\ket{m+1},\nonumber\\
\omega&=e^{2\pi\ii/N}, & ZX&=\omega XZ,
\label{eq:clock_shift}
\end{align}
with $Z^N=X^N=\id$. For $N=2$, $H$ reduces to the Hermitian transverse-field Ising chain; for $N>2$, it is generally non-Hermitian.

The Hamiltonian conserves the global $Z_N$ charge,
\begin{equation}
Q=\prod_{j=1}^{L}X_j,\qquad
Q\ket{\Psi_q}=\omega^q\ket{\Psi_q},
\label{eq:charge}
\end{equation}
dividing the Hilbert space into $N$ sectors ($q=0,1,\ldots,N-1$), each of dimension $D=N^{L-1}$. Homogeneous PBC also preserves lattice translation symmetry. For real $\lambda$ and $\mu$, spatial reflection combined with complex conjugation in the $Z$ basis relates the spectra of opposite charge sectors,
\begin{equation}
E\in\sigma(H_q)
\quad\Longrightarrow\quad
E^*\in\sigma(H_{-q}),
\label{eq:PT_pairing}
\end{equation}
so the spectrum in the neutral sector $q=0$ is symmetric about the real axis. Unless stated otherwise, we work in $q=0$ at $\lambda=1$ and denote the simple, real ground state (ordered by $\mathrm{Re}\,E$) with the subscript $0$.

Under OBC ($\mu=0$), the chain is integrable: its spectrum is built from independent free-parafermion modes, and it possesses mutually commuting conserved charges $\{J_m\}_{m=1}^L$ with $J_1=-H_{\mathrm{OBC}}$~\cite{Fendley2014}. Once the boundary bond $\mu$ is turned on, the OBC conserved charges are broken; at homogeneous PBC ($\mu=\lambda$), however, the periodic chain possesses its own family of mutually commuting conserved charges $\{I_m\}_{m=1}^L$ with $I_1=-H_{\mathrm{PBC}}$, although its spectrum no longer factorizes into state-independent free-parafermion modes~\cite{LiLiuBatchelor2026}.

\subsection{Directed hopping in configuration space}
\label{subsec:directed_hopping}

To obtain an intuitive physical picture of the NHSE in this spin chain, we work in the local $X$ eigenbasis,
\begin{align}
\ket{k}_X
&=\frac{1}{\sqrt N}
\sum_{m=0}^{N-1}\omega^{km}\ket{m},
\nonumber\\
X\ket{k}_X&=\omega^{-k}\ket{k}_X,\qquad
Z\ket{k}_X=\ket{k+1}_X.
\label{eq:Fourier}
\end{align}
In this basis, the field operators $X_j$ are diagonal, while each interaction $Z_j^\dagger Z_{j+1}$ shifts neighboring clock labels as $(k_j,k_{j+1})\to(k_j-1,k_{j+1}+1)$, acting as directed hopping between many-body configurations.

Within a fixed charge sector $q$, only $L-1$ labels $\bm k=(k_1,\ldots,k_{L-1})$ are independent, with the last label fixed by
\begin{equation}
k_L=-q-\sum_{j=1}^{L-1}k_j\pmod N.
\label{eq:charge_constraint}
\end{equation}
The sector Hamiltonian then takes the exact tight-binding form
\begin{align}
H_q\ket{\bm k;q}_X
={}&-d_q(\bm k)\ket{\bm k;q}_X
\nonumber\\
&-\lambda\sum_{j=1}^{L-1}
\ket{\bm k+\bm v_j;q}_X
-\mu\ket{\bm k+\bm v_L;q}_X,
\label{eq:graphH}
\end{align}
where the complex on-site potential is $V_q(\bm k)=-d_q(\bm k)$, with
\begin{equation}
d_q(\bm k)=\sum_{j=1}^{L-1}\omega^{-k_j}
+\omega^{q+\sum_{j=1}^{L-1}k_j},
\label{eq:potential}
\end{equation}
and the directed hopping vectors are
\begin{align}
\bm v_j&=-\bm e_j+\bm e_{j+1},
&&j=1,\ldots,L-2,\nonumber\\
\bm v_{L-1}&=-\bm e_{L-1},
&\bm v_L&=\bm e_1,
\label{eq:hopping_vectors}
\end{align}
with $\bm e_j$ the $j$th coordinate unit vector. The first $L-1$ hopping vectors come from bulk interactions, while $\bm v_L=\bm e_1$ represents the physical closing bond $B=Z_L^\dagger Z_1$.

\begin{figure}[t]
\centering
\plotfile[width=\columnwidth]{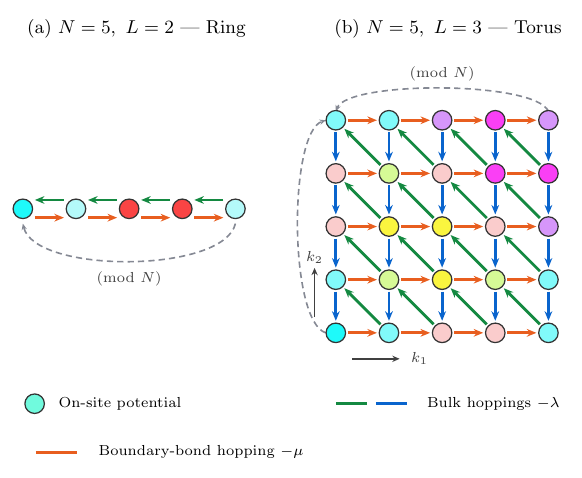}
\caption{Configuration-space representation of the Baxter--Fendley $Z_N$ spin chain for $N=5$ and $q=0$. (a) For $L=2$, configuration space forms a one-dimensional periodic ring of $N$ sites. (b) For $L=3$, it forms a two-dimensional discrete torus of $N^2$ sites. Each vertex represents a many-body configuration $\bm k$, colored by its complex on-site potential $V_q(\bm k)=-d_q(\bm k)$. Green and blue arrows represent bulk hopping ($-\lambda$), while orange arrows represent hopping induced by the physical closing bond ($-\mu$). Dashed gray curves indicate periodic identification modulo $N$; wraparound hopping arrows are omitted for clarity.}
\label{fig:fig1}
\end{figure}

This configuration-space lattice has two notable geometric features (Fig.~\ref{fig:fig1}). First, because every $k_j$ is defined modulo $N$, it is a discrete $(L-1)$-dimensional periodic torus with no geometric boundary, even under physical OBC. Increasing $N$ enlarges the linear size along each periodic direction, whereas increasing $L$ raises the dimension of configuration space. Second, closing the physical chain does not alter this periodic geometry: under OBC, the $L-1$ bulk hoppings already connect all configurations on the torus, and turning on $\mu$ simply adds one more directed hopping channel $\bm v_L=\bm e_1$ at every vertex.

From this lattice picture, the physical origin of the boundaryless NHSE is transparent. Writing Eq.~\eqref{eq:graphH} as $H_q=V_q+T$, both the diagonal potential $V_q$ and the torus shift operator $T$ are individually normal ($[V_q,V_q^\dagger]=[T,T^\dagger]=0$), so neither term alone can break eigenvector orthogonality. Their sum, however, is generically non-normal:
\begin{equation}
[H_q,H_q^\dagger]=[V_q,T^\dagger]+[T,V_q^\dagger].
\label{eq:nonnormality_commutator}
\end{equation}
When $V_q(\bm k)$ is nonuniform, this commutator is generally nonzero. As an eigenstate propagates along directed hopping paths through the nonuniform complex potential, it experiences spatially varying amplification and attenuation, enabling eigenstates to accumulate in preferred subregions of a closed torus without requiring any geometric edge.

\subsection{Condition number, boundary response, and configuration inversion}
\label{subsec:diagnostics}

For a diagonalizable non-normal Hamiltonian, the right and left eigenvectors form a biorthogonal basis,
\begin{equation}
H\ket{\psi_n^{\mathrm R}}=E_n\ket{\psi_n^{\mathrm R}},
\qquad
H^\dagger\ket{\psi_n^{\mathrm L}}=E_n^*\ket{\psi_n^{\mathrm L}},
\label{eq:left_right}
\end{equation}
whereas at EPs eigenvalues and eigenvectors coalesce and $H$ becomes defective. When visualizing low-dimensional states, we use separately normalized eigenvectors to define the configuration-space probabilities
\begin{equation}
P_n^{\mathrm{R,L}}(\bm k)=|\psi_n^{\mathrm{R,L}}(\bm k)|^2,
\qquad \sum_{\bm k}P_n^{\mathrm{R,L}}(\bm k)=1.
\label{eq:probabilities}
\end{equation}
For larger $L$, where high-dimensional wavefunctions are difficult to inspect and localization depends on the basis, we quantify eigenvector nonorthogonality using the eigenvalue condition number~\cite{TrefethenEmbree,NonnormalHamOkuma,pseudospectraNHSE1,pseudospectraNHSE2,GeneralCriterionNHSE,FengNumericalInstability,scalingpseudospectraexponentiallysensitive,naves2026levelrepulsionfailsnonnormality}. For any simple eigenvalue $E_n$,
\begin{equation}
\kappa_n=
\frac{\|\psi_n^{\mathrm L}\|\|\psi_n^{\mathrm R}\|}
{|\langle\psi_n^{\mathrm L}|\psi_n^{\mathrm R}\rangle|}
\geq 1,
\label{eq:eigenvalue_condition}
\end{equation}
where all vector and operator norms refer to the 2-norm (we also denote $\kappa_{\max}=\max_n\kappa_n$ over simple levels). For any normal matrix, $\kappa_n=1$, whereas $\kappa_n\gg 1$ signals strong left--right nonorthogonality. Crucially, $\kappa_n$ is invariant under unitary basis transformations, resolves individual eigenstates (enabling MPS evaluation without constructing the full eigenbasis), and directly governs linear spectral sensitivity.

Indeed, under a small perturbation $H\to H+\epsilon W$, first-order perturbation theory gives
\begin{equation}
E_n^{(1)}
=\frac{\langle\psi_n^{\mathrm L}|W|\psi_n^{\mathrm R}\rangle}
{\langle\psi_n^{\mathrm L}|\psi_n^{\mathrm R}\rangle},
\qquad
\kappa_n=\max_{\|W\|_2=1}|E_n^{(1)}|,
\label{eq:optimal_response}
\end{equation}
where the maximum is over arbitrary unit-norm complex perturbations. To isolate the actual sensitivity to the physical closing bond $B=Z_L^\dagger Z_1$ ($\|B\|_2=1$), we define the boundary eigenvalue condition number
\begin{equation}
\kappa_{\mathrm b,n}
\equiv
\left|\frac{\partial E_n}{\partial\mu}\right|
=\frac{|\langle\psi_n^{\mathrm L}|B|\psi_n^{\mathrm R}\rangle|}
{|\langle\psi_n^{\mathrm L}|\psi_n^{\mathrm R}\rangle|}
=\eta_{\mathrm b,n}\kappa_n,
\label{eq:boundary_response_definition}
\end{equation}
with the bond-alignment factor
\begin{equation}
\eta_{\mathrm b,n}
=\frac{|\langle\psi_n^{\mathrm L}|B|\psi_n^{\mathrm R}\rangle|}
{\|\psi_n^{\mathrm L}\|\|\psi_n^{\mathrm R}\|}
\leq 1.
\label{eq:alignment_factor}
\end{equation}
This factorization separates two distinct effects: $\kappa_n$ measures the maximum spectral amplification available from nonorthogonality, whereas $\eta_{\mathrm b,n}$ measures how effectively the physical closing bond couples the left and right eigenstates.

Finally, in $q=0$ (and $q=N/2$ for even $N$) with real couplings, configuration inversion $\mathcal R\ket{\bm k;q}_X=\ket{-\bm k;q}_X$ obeys
\begin{equation}
H_q^\dagger=\mathcal R H_q\mathcal R,
\qquad \mathcal R^2=\id.
\label{eq:configuration_inversion}
\end{equation}
For any simple real eigenvalue, this allows choosing $\ket{\psi_n^{\mathrm L}}=\mathcal R\ket{\psi_n^{\mathrm R}}$, so that $\kappa_n=1/|\langle\psi_n^{\mathrm R}|\mathcal R|\psi_n^{\mathrm R}\rangle|$ for normalized eigenvectors. Whenever a right wave packet localizes away from inversion-symmetric points of the torus, its left partner localizes at $-\bm k$, directly suppressing their overlap and driving a large $\kappa_n$.

\section{Non-Hermitian skin effect on boundaryless lattices}
\label{sec:skin}

We first examine the two simplest chain lengths, $L=2$ and $L=3$, as the clock dimension $N$ increases. In these low-dimensional cases, configuration space forms a one-dimensional ring or a two-dimensional torus, allowing us to visualize the boundaryless NHSE directly and isolate its two key mechanisms: spatially varying amplification with basis duality on a ring ($L=2$), and multichannel path interference governed by integrability on a torus ($L=3$). For both geometries, we also establish the large-$N$ scalings of $\kappa_n$ and $\kappa_{\mathrm b,n}$.

\subsection{Directed ring for $L=2$}
\label{subsec:ring}

For $L=2$, each charge sector maps onto a periodic ring of $N$ configurations. In the $X$ basis, Eq.~\eqref{eq:graphH} reads
\begin{align}
H_q\ket{k;q}_X
={}&-(\omega^{-k}+\omega^{q+k})\ket{k;q}_X\nonumber\\
&-\lambda\ket{k-1;q}_X-\mu\ket{k+1;q}_X.
\label{eq:ring_hamiltonian}
\end{align}
Physical OBC ($\mu=0$) gives unidirectional hopping around the ring, whereas closing the chain ($\mu=\lambda$) adds hopping in the reverse direction without altering the periodic geometry.

\begin{figure}[t]
\centering
\plotfile[width=\columnwidth]{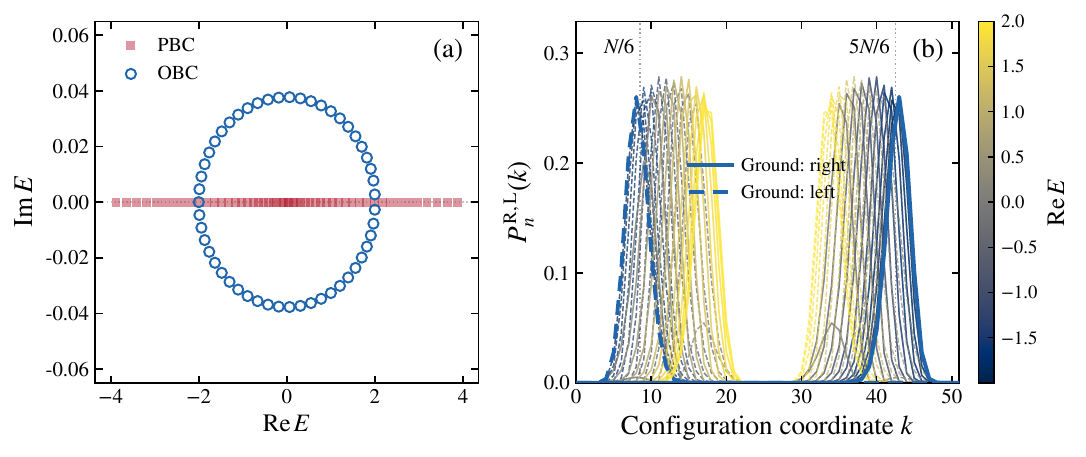}
\caption{Spectrum and boundaryless NHSE on the $L=2$ configuration-space ring for $N=51$, $q=0$, and $\lambda=1$.
(a) OBC (blue open circles) and PBC (red squares) energy spectra. Under OBC the eigenvalues form a closed complex loop with only a single real level (the ground state), whereas under PBC the Hamiltonian in this sector is Hermitian and the spectrum is purely real.
(b) Right (solid) and left (dashed) OBC eigenstate probabilities $P_n^{\mathrm{R,L}}(k)$, colored by $\mathrm{Re}\,E_n$. The ground-state profiles are highlighted in blue, and vertical dotted lines mark the analytical packet centers $k_{\mathrm L}=N/6$ and $k_{\mathrm R}=5N/6$.}
\label{fig:fig2}
\end{figure}

We first consider OBC ($\mu=0$) at $\lambda=1$ in the neutral sector $q=0$, where
\begin{equation}
H_0^{\mathrm{OBC}}\ket{k}_X
=-2\cos\!\left(\frac{2\pi k}{N}\right)\ket{k}_X-\ket{k-1}_X.
\label{eq:ring_obc}
\end{equation}
The right-eigenstate equation reduces to the one-step recurrence
\begin{align}
\psi^{\mathrm R}(k+1)&=g_k(E)\psi^{\mathrm R}(k),\nonumber\\
g_k(E)&=-E-2\cos\!\left(\frac{2\pi k}{N}\right).
\label{eq:ring_recursion}
\end{align}
Here $|g_k(E)|$ is the spatial amplitude ratio between neighboring configurations. Periodicity around the ring requires $\prod_{k=0}^{N-1}g_k(E)=1$, which allows $|g_k(E)|>1$ along one arc of the ring and $|g_k(E)|<1$ along the remaining arc. Wavefunctions therefore grow across one part of the ring and decay across the other, causing many right eigenstates to accumulate in overlapping subregions despite the absence of any geometric edge [Fig.~\ref{fig:fig2}(b)].

The OBC ground state provides an exact analytical illustration (Appendix~\ref{app:ring_obc}). Its energy is
\begin{equation}
E_0=-2\cosh\!\left(\frac{a}{N}\right),
\qquad a=\operatorname{arcosh}\!\left(\frac32\right).
\label{eq:ring_energy}
\end{equation}
At large $N$, $g_k(E_0)$ is positive and approaches $4\sin^2(\pi k/N)$, exceeding unity for $N/6<k<5N/6$ and falling below unity elsewhere. Consequently, the right ground-state wave packet peaks near $k_{\mathrm R}$, while configuration inversion~\eqref{eq:configuration_inversion} places its left partner near $k_{\mathrm L}$:
\begin{equation}
k_{\mathrm R}\simeq\frac{5N}{6},\qquad
k_{\mathrm L}\simeq\frac{N}{6}.
\label{eq:ring_centers}
\end{equation}
Because the packet widths scale only as $O(\sqrt N)$ whereas their separation scales linearly as $O(N)$, the left--right overlap decreases exponentially with $N$. Integrating the spatial gain across the amplification arc yields
\begin{align}
\ln\kappa_0^{\mathrm{OBC}}&=c_{\mathrm O}N+O(\ln N),\nonumber\\
c_{\mathrm O}&=\int_{1/6}^{5/6}\!dx\,\ln[4\sin^2(\pi x)]\simeq0.6461.
\label{eq:ring_scaling}
\end{align}
Excited OBC states display similar left--right packet separation and exponential growth of $\kappa_n^{\mathrm{OBC}}$ with $N$ [Fig.~\ref{fig:fig2}(b)].

In contrast, the linear response of the OBC ground state to the physical closing bond $B=Z_2^\dagger Z_1$ grows only polynomially with $N$ (Appendix~\ref{app:bond_large_n}):
\begin{equation}
\kappa_{\mathrm b,0}^{\mathrm{OBC}}
\sim C_2 N^2,\qquad
C_2=\frac{1}{2(\pi^2+a^2)}\simeq 0.04631.
\label{eq:ring_bond}
\end{equation}
Physically, $B$ shifts configurations by only one step and cannot bridge the $O(N)$ separation between the left and right wave packets. The alignment factor $\eta_{\mathrm b,0}=\kappa_{\mathrm b,0}/\kappa_0$ is therefore exponentially suppressed, leaving only a polynomial $N^2$ enhancement governed by the wavefunction tails near $k=0$.

The PBC ring ($\mu=\lambda=1$) provides two instructive contrasts. First, in the neutral sector $q=0$,
\begin{equation}
H_0^{\mathrm{PBC}}\ket{k}_X
=-2\cos\!\left(\frac{2\pi k}{N}\right)\ket{k}_X
-\ket{k-1}_X-\ket{k+1}_X,
\label{eq:ring_pbc_zero}
\end{equation}
the hopping is reciprocal and the potential is real, so $H_0^{\mathrm{PBC}}$ is Hermitian and its spectrum collapses onto the real axis [Fig.~\ref{fig:fig2}(a)]. Although its ground state forms a localized Gaussian packet of width $O(\sqrt N)$ centered near $k=0$, its left and right eigenvectors coincide identically, giving $\kappa_0^{\mathrm{PBC}}=1$ for every $N$ and $\kappa_{\mathrm b,0}^{\mathrm{PBC}}\to 1$ at large $N$. Wavefunction concentration alone therefore does not imply non-normality or the NHSE.

\begin{figure}[t]
\centering
\plotfile[width=\columnwidth]{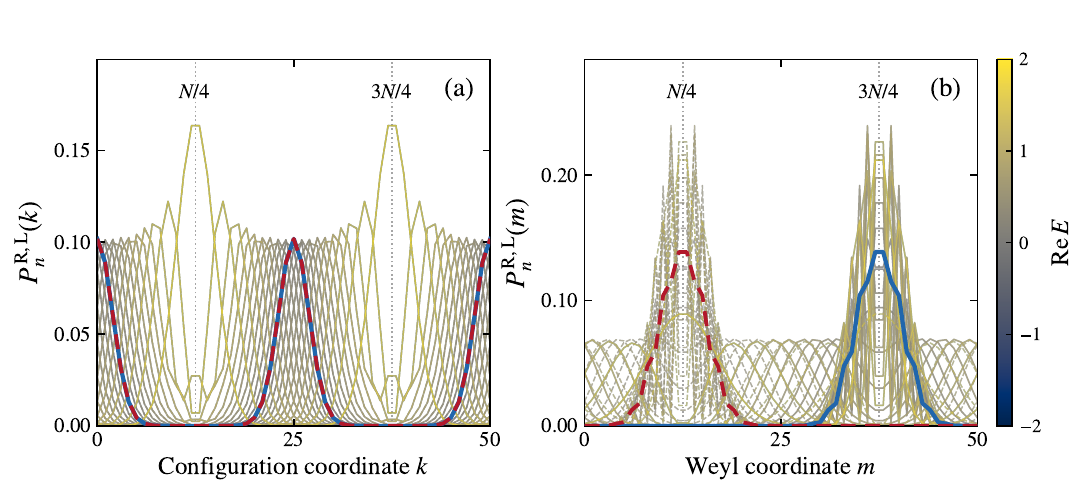}
\caption{Basis duality of boundaryless PBC NHSE for $L=2$, $N=50$, $q=N/2$, and $\lambda=\mu=1$. Right (solid) and left (dashed) eigenstate probabilities are shown in (a) the $X$ basis and (b) the unitarily equivalent Weyl basis. Thin curves are colored by $\mathrm{Re}\,E_n$, while thick blue (right) and dashed red (left) curves highlight the same strongly non-normal near-zero mode in both panels. Its left and right probabilities coincide in the $X$ basis but form separated packets near $m=3N/4$ and $N/4$ (vertical dotted lines) in the Weyl basis.}
\label{fig:fig3}
\end{figure}

Second, nontrivial charge sectors demonstrate that boundaryless NHSE also emerges under physical PBC. For even $N$ at $q=N/2$ and $\lambda=\mu=1$,
\begin{equation}
H_{N/2}^{\mathrm{PBC}}\ket{k}_X
=2\ii\sin\!\left(\frac{2\pi k}{N}\right)\ket{k}_X
-\ket{k-1}_X-\ket{k+1}_X.
\label{eq:ring_pbc_half}
\end{equation}
Although hopping is reciprocal in the $X$ basis, the imaginary on-site potential makes $H_{N/2}^{\mathrm{PBC}}$ complex symmetric and non-normal. For a simple eigenvalue, $\psi_n^{\mathrm L}(k)=[\psi_n^{\mathrm R}(k)]^*$, so left and right probability distributions coincide in the $X$ basis [Fig.~\ref{fig:fig3}(a)], while phase cancellation suppresses their overlap:
\begin{equation}
\kappa_n=\frac{1}{\left|\sum_k[\psi_n^{\mathrm R}(k)]^2\right|}.
\label{eq:ring_phase_cancellation}
\end{equation}

This hidden nonorthogonality becomes spatially transparent in the unitary Weyl basis,
\begin{equation}
\ket{m}_W=\frac{1}{\sqrt N}\sum_{k=0}^{N-1}
\exp\!\left(\frac{2\pi\ii mk}{N}
-\frac{\ii\pi k^2}{N}\right)\ket{k}_X,
\label{eq:weyl_basis}
\end{equation}
which eliminates the imaginary on-site potential and transforms Eq.~\eqref{eq:ring_pbc_half} into a pure nearest-neighbor hopping ring with position-dependent asymmetric amplitudes (Appendix~\ref{app:ring_pbc}),
\begin{align}
\widetilde H&=\sum_{m=0}^{N-1}\bigl(
a_m\ket{m+1}_W{}_W\bra{m}
+b_m\ket{m}_W{}_W\bra{m+1}\bigr),\nonumber\\
a_m&=1-e^{-\ii\theta_m},\qquad
b_m=-1-e^{\ii\theta_m},
\label{eq:weyl_hamiltonian}
\end{align}
where $\theta_m=2\pi(m+1/2)/N$. The local hopping asymmetry obeys
\begin{equation}
\left|\frac{a_m}{b_m}\right|
=\left|\tan\!\frac{\pi(m+1/2)}{N}\right|,
\label{eq:weyl_asymmetry}
\end{equation}
whose preferred hopping direction reverses near $m=N/4$ and $3N/4$. Consequently, highly ill-conditioned real near-zero modes separate into right and left packets around $3N/4$ and $N/4$, satisfying $P_n^{\mathrm L}(m)=P_n^{\mathrm R}(-m)$ modulo $N$ [Fig.~\ref{fig:fig3}(b)]. Accumulating this hopping asymmetry between the two packet regions along $N=4M+2$ gives the semiclassical estimate (Appendix~\ref{app:ring_pbc})
\begin{equation}
\ln\kappa_{\mathrm{nz}}^{\mathrm{PBC}}\simeq\gamma_{\mathrm P}N,
\qquad \gamma_{\mathrm P}=\frac{G}{\pi}\simeq 0.2916,
\label{eq:weyl_scaling}
\end{equation}
where $G$ is Catalan's constant (consistent with direct diagonalization: $\kappa_{\mathrm{nz}}^{\mathrm{PBC}}\simeq 21.55$ at $N=18$ and $1.38\times 10^5$ at $N=50$). Because the Weyl basis is unitarily equivalent to the $X$ basis, $\kappa_n$ is identical in both: phase cancellation in one basis becomes spatial separation in the other. Both physical OBC and PBC can therefore host boundaryless NHSE on a closed ring.

\subsection{Directed torus for $L=3$}
\label{subsec:torus}

For $L=3$, configuration space becomes a two-dimensional periodic torus. Setting $\lambda=1$ in the $q=0$ sector, the $X$-basis Hamiltonian is
\begin{align}
H(\mu)\ket{k_1,k_2}
={}&-d_0(\bm k)\ket{k_1,k_2}
-\ket{k_1-1,k_2+1}\nonumber\\
&-\ket{k_1,k_2-1}-\mu\ket{k_1+1,k_2},
\label{eq:torus_hamiltonian}
\end{align}
with $d_0(\bm k)=\omega^{-k_1}+\omega^{-k_2}+\omega^{k_1+k_2}$ and coordinates modulo $N$. Under OBC ($\mu=0$), each configuration receives amplitude from two incoming hopping channels,
\begin{align}
[E+d_0(\bm k)]\psi^{\mathrm R}(k_1,k_2)
={}&-\psi^{\mathrm R}(k_1+1,k_2-1)\nonumber\\
&-\psi^{\mathrm R}(k_1,k_2+1).
\label{eq:torus_recursion}
\end{align}
Unlike the single-channel recurrence~\eqref{eq:ring_recursion} for $L=2$, Eq.~\eqref{eq:torus_recursion} determines only the sum of two incoming amplitudes, so interference between competing paths becomes essential.

\begin{figure}[t]
\centering
\plotfile[width=\columnwidth]{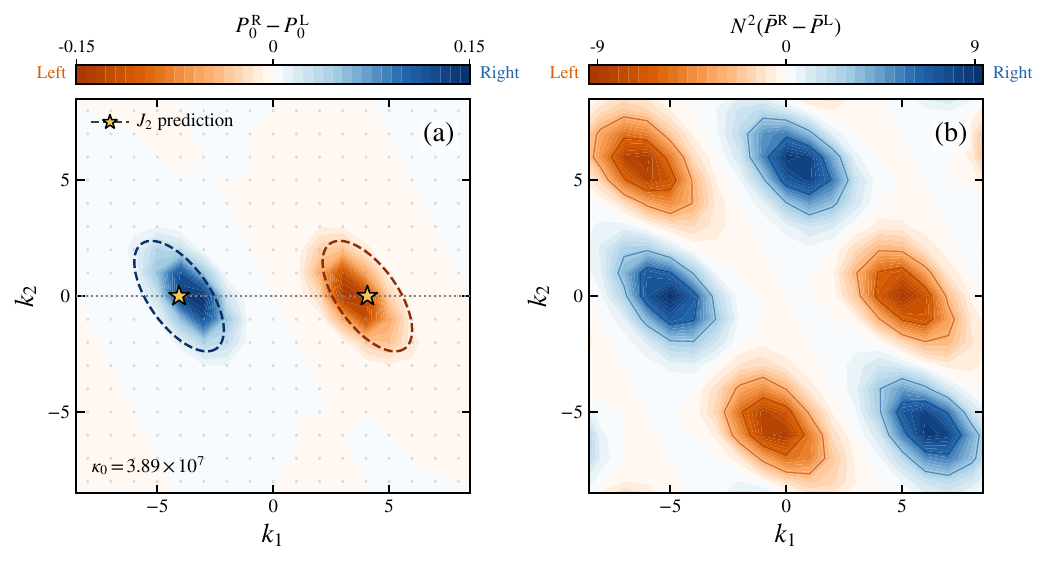}
\caption{Boundaryless NHSE on the $L=3$ configuration-space torus for OBC, $N=17$, $q=0$, and $\lambda=1$, plotted in centered periodic coordinates $(k_1,k_2)$.
(a) Ground-state probability difference $P_0^{\mathrm R}-P_0^{\mathrm L}$ (blue: right; orange: left), showing well-separated packets with $\kappa_0^{\mathrm{OBC}}=3.89\times 10^7$. Gold stars and dashed ellipses mark the analytical packet centers and $1/e^2$ Gaussian contours predicted from the OBC conserved charge $J_2$; the horizontal dotted line indicates $k_2=0$.
(b) Ensemble-averaged probability difference $N^2(\bar{P}^{\mathrm R}-\bar{P}^{\mathrm L})$ over all $N^2$ eigenstates, normalized by the uniform normal-matrix baseline $1/N^2$. The alternating three blue and three orange lobes demonstrate extensive left--right accumulation across the entire spectrum.}
\label{fig:fig4}
\end{figure}

Here, the integrability of the OBC chain directly resolves this two-channel interference. Alongside $J_1=-H_{\mathrm{OBC}}$, the OBC chain possesses a second commuting conserved charge $J_2$~\cite{Fendley2014}, supplying an independent exact relation for the same two incoming amplitudes (Appendix~\ref{app:torus_joint}). Solving the joint eigenvalue equations of $H_{\mathrm{OBC}}$ and $J_2$ semiclassically at large $N$ along $k_2=0$ ($u=k_1/N$) gives the local ground-state amplitude ratios $p_1=\psi_0^{\mathrm R}(k_1+1,-1)/\psi_0^{\mathrm R}(k_1,0)$ and $p_2=\psi_0^{\mathrm R}(k_1,1)/\psi_0^{\mathrm R}(k_1,0)$ (Appendix~\ref{app:torus_wkb}):
\begin{align}
p_{1,2}&\simeq G(u)e^{\pm\ii\theta(u)},\nonumber\\
G(u)&=4\sin(\pi u)\cos\!\left(\frac{\pi u}{2}\right),\qquad
\theta(u)=\frac{\pi}{2}(1-u).
\label{eq:torus_channels}
\end{align}
Physically, the two incoming paths share the same local gain factor $G(u)$ but carry opposite phases $\pm\theta(u)$, interfering coherently as $2G\cos\theta$ in Eq.~\eqref{eq:torus_recursion}. Advancing one step along $k_1$ at fixed $k_2$ combines both channels via their product:
\begin{equation}
\left|\frac{\psi_0^{\mathrm R}(k_1+1,0)}{\psi_0^{\mathrm R}(k_1,0)}\right|
\simeq |p_1 p_2|=G^2(u).
\label{eq:torus_gain}
\end{equation}
The right ground-state amplitude therefore grows where $G(u)>1$ and decays where $G(u)<1$, peaking at the larger root of $G(u)=1$:
\begin{equation}
\bm k_{\mathrm R}\simeq N(u_{\mathrm R},0),\qquad
u_{\mathrm R}\simeq 0.7611.
\label{eq:torus_center}
\end{equation}
The local semiclassical Hessian at $(u_{\mathrm R},0)$ is negative definite with nonzero off-diagonal entries (Appendix~\ref{app:torus_wkb}), confining the right ground state to a tilted two-dimensional Gaussian packet of width $O(\sqrt N)$. By configuration inversion~\eqref{eq:configuration_inversion}, the left ground state localizes around $\bm k_{\mathrm L}\simeq N(1-u_{\mathrm R},0)$, separated from $\bm k_{\mathrm R}$ by an $O(N)$ distance on the periodic torus. Figure~\ref{fig:fig4}(a) illustrates this spatial separation via the ground-state probability difference $P_0^{\mathrm R}(\bm k)-P_0^{\mathrm L}(\bm k)$ at $N=17$ in centered periodic coordinates $k_{1,2}\in[-(N-1)/2,(N-1)/2]$. In these coordinates, the right (blue) and left (orange) packets sit symmetrically on opposite sides of the origin along $k_2=0$, centered at $k_1\simeq(u_{\mathrm R}-1)N$ and $(1-u_{\mathrm R})N$, respectively. Both the predicted centers (gold stars) and the tilted $1/e^2$ Gaussian contours (dashed ellipses) derived from $J_2$ match exact diagonalization closely, directly accounting for $\kappa_0^{\mathrm{OBC}}=3.89\times 10^7$. This integrability-assisted construction generalizes naturally to higher-dimensional configuration-space tori: for an $L$-site OBC chain, the $L-1$ incoming bulk hopping channels at each vertex can in principle be fixed by the commuting conserved charges $J_1,\ldots,J_{L-1}$ (for example, $J_1$, $J_2$, and $J_3$ on the three-dimensional torus for $L=4$) to determine the semiclassical skin-mode packet.

To establish that this behavior is a genuine NHSE---involving macroscopic accumulation of an extensive fraction of the spectrum rather than isolated-state localization---Fig.~\ref{fig:fig4}(b) shows the ensemble-averaged probability difference $N^2[\bar{P}^{\mathrm R}(\bm k)-\bar{P}^{\mathrm L}(\bm k)]$, where $\bar{P}^{\mathrm{R,L}}(\bm k)=N^{-2}\sum_{n=1}^{N^2}P_n^{\mathrm{R,L}}(\bm k)$ with each right and left eigenvector separately normalized as in Eq.~\eqref{eq:probabilities}. For any normal Hamiltonian (even with localized eigenstates), eigenvector unitarity enforces $\bar{P}^{\mathrm R}=\bar{P}^{\mathrm L}\equiv 1/N^2$, so this map vanishes identically; likewise, localization of only a few states would be washed out by the $1/N^2$ average. Instead, Fig.~\ref{fig:fig4}(b) exhibits a pronounced alternating pattern of three right-dominated (blue) and three left-dominated (orange) lobes across the torus. This demonstrates that an extensive number of right eigenstates collectively pile up in preferred subregions while their left counterparts accumulate in distinct regions, providing direct evidence of a boundaryless NHSE throughout the spectrum. This approximate threefold pattern at $N=17$ originates from a large-$N$ spectral symmetry: shifting $(k_1,k_2)\to(k_1\pm N/3,k_2\pm N/3)$ multiplies the on-site potential $d_0(\bm k)$ by $e^{\mp 2\pi\ii/3}$, and combining this shift with a position-dependent phase transformation multiplies the entire Hamiltonian by the same phase---becoming the exact threefold spectral rotation of Appendix~\ref{app:ep_jordan} whenever $N$ is a multiple of $3$. Accordingly, the three rotationally related branches of the OBC complex spectrum accumulate in three well-separated right regions interleaved with their inverted left counterparts.

We now summarize the large-$N$ scalings of $\kappa_n$ and $\kappa_{\mathrm b,n}$ for $L=3$. Under OBC, global matching of the semiclassical ground-state envelope yields an exponential scaling (Appendix~\ref{app:torus_overlap}),
\begin{equation}
\ln\kappa_0^{\mathrm{OBC}}\simeq c_{\mathrm{sc}}N,\qquad
c_{\mathrm{sc}}\simeq 1.8351,
\label{eq:torus_scaling}
\end{equation}
consistent with numerical diagonalization ($\kappa_0^{\mathrm{OBC}}\simeq 3.89\times 10^7$ at $N=17$ and $6.63\times 10^{11}$ at $N=23$). Although excited OBC states can attain even larger condition numbers, the integrable spectral-projector construction (Appendix~\ref{app:projector_bounds}) proves a uniform upper bound $\ln\kappa_{\max}^{\mathrm{OBC}}=O(N)$ for all simple eigenvalues at any fixed $L$. By contrast, the linear closing-bond response of the OBC ground state grows only polynomially at fixed $L$ (Appendix~\ref{app:bond_large_n}):
\begin{equation}
\kappa_{\mathrm b,0}^{\mathrm{OBC}}\sim C_3 N^4,\qquad
C_3\simeq 0.003014,
\label{eq:torus_bond}
\end{equation}
and more generally $\kappa_{\mathrm b,0}^{\mathrm{OBC}}\sim C_L N^{2L-2}$ for fixed $L$. As in the $L=2$ ring, the contrast between exponential $\kappa_0^{\mathrm{OBC}}$ and polynomial $\kappa_{\mathrm b,0}^{\mathrm{OBC}}$ implies an exponentially small bond-alignment factor $\eta_{\mathrm b,0}$.

Under homogeneous PBC ($\mu=1$), the closing bond adds the third hopping vector $(1,0)$. Because the three hopping displacements sum to zero,
\begin{equation}
(-1,1)+(0,-1)+(1,0)=\bm 0,
\label{eq:torus_drift}
\end{equation}
their first-order spatial derivatives cancel for smooth envelopes near $\bm k=\bm 0$. The low-energy effective Hamiltonian therefore reduces to a Hermitian two-dimensional harmonic oscillator (Appendix~\ref{app:torus_pbc}), whose real Gaussian ground state has width $O(\sqrt N)$ and coinciding left and right wavefunctions, yielding
\begin{equation}
\kappa_0^{\mathrm{PBC}}\longrightarrow 1,
\qquad
\kappa_{\mathrm b,0}^{\mathrm{PBC}}\longrightarrow 1
\qquad (N\to\infty,\ L=3),
\label{eq:torus_pbc_limit}
\end{equation}
in agreement with direct diagonalization ($\kappa_0^{\mathrm{PBC}}\simeq 1.0038$ at $N=20$ and $1.0018$ at $N=40$). However, this emergent Hermiticity applies only near the ground state. Excited PBC eigenstates in the same $q=0$ sector (resolving translation symmetry and excluding EPs) remain strongly nonorthogonal, with $\kappa_{\max}^{\mathrm{PBC}}\simeq 4.64\times 10^2$ at $N=20$ and $7.54\times 10^5$ at $N=40$, growing rapidly with $N$. Thus, both OBC and PBC exhibit boundaryless NHSE on the two-dimensional torus, while the ground-state response to the single closing bond stays polynomial (OBC) or $O(1)$ (PBC) at fixed $L$. In the next section, we show that increasing the chain length $L$ at fixed $N$ leads to a qualitatively different many-body enhancement.

\section{Many-body enhancement and boundary sensitivity}
\label{sec:manybody}

We now turn to longer chains by increasing $L$ at fixed clock dimension $N$ (focusing on $N=3$ and $\lambda=1$ for numerical illustrations). Unlike the large-$N$ limit of Sec.~\ref{sec:skin}, which enlarges the linear size of a low-dimensional torus, increasing $L$ raises the dimension $d=L-1$ of configuration space so that the number of configurations grows exponentially as $N^{L-1}$.

\subsection{Ground-state nonorthogonality from commuting-charge MPOs}
\label{subsec:manybody_kappa}

To access chain lengths up to $L=50$---well beyond exact diagonalization (ED)---we exploit commuting conserved charges under both OBC and PBC to construct ground-state wavefunctions via MPS.

Under OBC, the commuting conserved charges~\cite{Fendley2014} form the generating operator
\begin{equation}
\mathcal C(t)=\sum_{m=0}^{L}t^m J_m,\qquad J_0=\id,\quad J_1=-H_{\mathrm{OBC}}.
\label{eq:charge_generating}
\end{equation}
On an eigenstate with free-parafermion occupations $\bm p=(p_1,\ldots,p_L)$, its eigenvalue at $\lambda=1$ is
\begin{equation}
c_{\bm p}(t)=\prod_{j=1}^{L}
\left(1+t\epsilon_j\omega^{p_j}\right),\qquad
\epsilon_j=\left[2\cos\frac{\pi j}{2L+1}\right]^{2/N}.
\label{eq:charge_eigenvalues}
\end{equation}
The ground state corresponds to $p_j=0$ for all $j$. Evaluating $\mathcal C(t)$ at the zeros $t_{js}=-\omega^s/\epsilon_j$ ($s=1,\ldots,N-1$) annihilates every excited occupation pattern while leaving the ground state invariant, yielding the exact OBC ground-state spectral projector:
\begin{equation}
\Pi_0=\prod_{j=1}^{L}\prod_{s=1}^{N-1}
\frac{\mathcal C(t_{js})}{c_{\bm 0}(t_{js})}
=\frac{\ket{\psi_0^{\mathrm R}}\bra{\psi_0^{\mathrm L}}}
{\langle\psi_0^{\mathrm L}|\psi_0^{\mathrm R}\rangle},
\qquad
\|\Pi_0\|_2=\kappa_0^{\mathrm{OBC}}.
\label{eq:ground_projector}
\end{equation}
Thus, the ground-state condition number equals the operator norm of this nonorthogonal spectral projector built directly from the conserved charges.

Crucially, $\Pi_0$ need not be assembled as a dense matrix. Each factor $\mathcal C(t)$ admits an exact bond-dimension-two matrix product operator (MPO) representation (Appendix~\ref{app:mpo}),
\begin{align}
\mathcal C(t)&=(1\ \ 0)\,W_1(t)\cdots W_L(t)
\begin{pmatrix}1\\0\end{pmatrix},\nonumber\\
W_j(t)&=\begin{pmatrix}
\id+tX_j&tZ_j^\dagger\\
Z_j&t\id
\end{pmatrix}.
\label{eq:charge_MPO}
\end{align}
Applying the $L(N-1)$ factors sequentially to the product state $\ket{\Omega}=\ket{0}_X^{\otimes L}$ and compressing the resulting MPS to bond dimension $\chi$ after each step yields $\ket v=\Pi_0\ket\Omega\propto\ket{\psi_0^{\mathrm R}}$. Using configuration inversion $H_0^\dagger=\mathcal R H_0\mathcal R$ together with $\mathcal R\ket\Omega=\ket\Omega$, we obtain two MPS estimators for $\kappa_0^{\mathrm{OBC}}$ (Appendix~\ref{app:mps_checks}):
\begin{equation}
\kappa_0^{\mathrm{OBC}}
=\frac{\langle v|v\rangle}{|\langle v|\mathcal R|v\rangle|}
=\frac{\langle v|v\rangle}{|\langle\Omega|v\rangle|}.
\label{eq:MPS_kappa}
\end{equation}
Benchmarked against left--right ED at $N=3$, $L=12$, and $\chi=128$, this method reproduces $\kappa_0^{\mathrm{OBC}}$ to within $2\times 10^{-13}$.

Under homogeneous PBC ($\lambda=\mu=1$), closing the chain breaks the OBC free-parafermion structure, so eigenstates no longer share a common set of state-independent zeros that can be annihilated in $O(L)$ exact steps. Nevertheless, summing over mutually nonadjacent subsets of all $2L$ local terms on the periodic ring defines a commuting-charge generating operator $\mathcal C_{\mathrm{PBC}}(t)=\sum_{m=0}^L t^m I_m$ ($I_0=\id$, $I_1=-H_{\mathrm{PBC}}$) satisfying $[\mathcal C_{\mathrm{PBC}}(t),H_{\mathrm{PBC}}]=0$ for all $t$~\cite{LiLiuBatchelor2026}. Partitioning these subsets according to whether the closing bond $B=Z_L^\dagger Z_1$ is absent or present splits $\mathcal C_{\mathrm{PBC}}(t)$ into a sum of two bond-dimension-two open-chain MPOs (Appendix~\ref{app:mpo}). Iteratively applying $\mathcal C_{\mathrm{PBC}}(t)$ with positive real $t$ to $\ket\Omega$ provides a high-degree polynomial MPS filter completely free of Trotter time-step error, from which we evaluate $\kappa_0^{\mathrm{PBC}}=\langle v|v\rangle/|\langle v|\mathcal R|v\rangle|$ (Appendix~\ref{app:mps_checks}).

\begin{figure}[t]
\centering
\plotfile[width=\columnwidth]{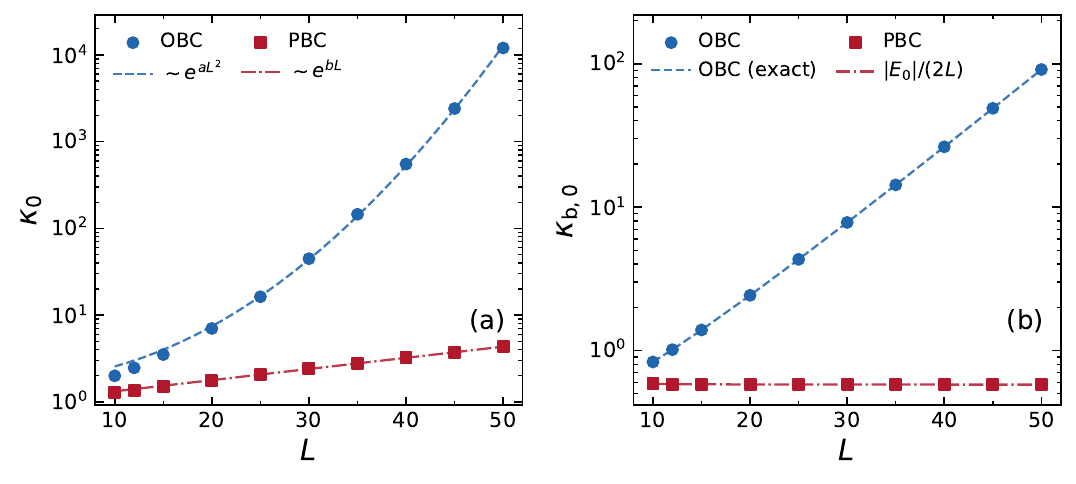}
\caption{Many-body enhancement and physical boundary response versus chain length $L$ ($10\le L\le 50$) for $N=3$, $q=0$, and $\lambda=1$ obtained from MPS calculations.
(a) Ground-state eigenvalue condition number $\kappa_0$ under OBC (blue circles) and PBC (red squares). The blue dashed curve and red dash-dotted line show finite-size fits over $20\le L\le 50$ to $\kappa_0^{\mathrm{OBC}}\sim e^{aL^2}$ ($a\simeq 0.0035$) and $\kappa_0^{\mathrm{PBC}}\sim e^{bL}$ ($b\simeq 0.0296$), respectively.
(b) Ground-state closing-bond response $\kappa_{\mathrm b,0}$ under OBC (blue circles) and PBC (red squares), benchmarked against the exact OBC integral~\eqref{eq:bond_integral} (blue dashed curve) and the exact PBC self-dual identity $|E_0|/(2L)$ (red dash-dotted line).
At $L=10$ and $12$, the MPS data agree with exact diagonalization to within symbol size (see Appendix~\ref{app:mps_checks} for convergence checks).}
\label{fig:fig5}
\end{figure}

Over $10\le L\le 50$ at $N=3$, our MPS calculations reveal a sharp contrast between OBC and PBC ground-state nonorthogonality [Fig.~\ref{fig:fig5}(a)], with finite-size fits over $20\le L\le 50$ obeying
\begin{equation}
\ln\kappa_0^{\mathrm{OBC}}\propto L^2,
\qquad
\ln\kappa_0^{\mathrm{PBC}}\propto L
\label{eq:manybody_trends}
\end{equation}
($a\simeq 0.0035$ and $b\simeq 0.0296$). Similarly, ED over accessible sizes indicates $\ln\kappa_{\max}\propto L^2$ under both OBC and PBC (excluding EPs), with $\kappa_{\max}$ typically occurring in excited states. Physically, $\ln\kappa_0$ measures how rapidly the right ground state becomes orthogonal to its inverted left partner as new configuration-space dimensions are added. Remarkably, bounding the operator norm of the exact projector~\eqref{eq:ground_projector} yields a rigorous analytical upper bound $\ln\kappa_0^{\mathrm{OBC}}=O(L^2)$ at fixed $N$ (Appendix~\ref{app:projector_bounds}), showing that the observed quadratic-exponential growth over $10\le L\le 50$ is consistent with the scaling of the integrable upper bound. Under PBC, the linear growth of $\ln\kappa_0^{\mathrm{PBC}}$ with $L$ arises naturally from a translationally invariant MPS description whenever the mixed inversion transfer matrix has a dominant eigenvalue of modulus $0<r<1$ (here $r=e^{-b}\simeq 0.9709$; see Appendix~\ref{app:pbc_transfer}). Note that this large-$L$ growth at fixed $N$ does not contradict Eq.~\eqref{eq:torus_pbc_limit}, which governs the opposite limit of large $N$ at fixed $L$.

\subsection{Response to physical boundary closure}
\label{subsec:manybody_boundary}

Does this rapid growth of $\kappa_0$ translate directly into strong sensitivity to the single physical closing bond $B=Z_L^\dagger Z_1$ in $H(\mu)=H_{\mathrm{OBC}}-\mu B$? By Eq.~\eqref{eq:boundary_response_definition}, the linear boundary response $\kappa_{\mathrm b,0}=\eta_{\mathrm b,0}\kappa_0$ depends critically on the bond-alignment factor $\eta_{\mathrm b,0}$.

For the OBC ground state, the exact free-parafermion solution gives~\cite{LiLiuBatchelor2026}
\begin{equation}
\kappa_{\mathrm b,0}^{\mathrm{OBC}}
=\frac{\sin(\pi/N)}{\pi}
\int_{-\infty}^{\infty}
\frac{dx}{\displaystyle\prod_{j=1}^{L}
|x-\epsilon_j e^{\ii\pi/N}|^2},
\label{eq:bond_integral}
\end{equation}
with $\epsilon_j$ from Eq.~\eqref{eq:charge_eigenvalues}. Unlike the fixed-$L$ power law in $N$, increasing $L$ at fixed $N>2$ adds a new factor to the denominator for each spin. A saddle-point analysis yields exponential growth with $L$ (Appendix~\ref{app:bond_large_l}):
\begin{align}
\ln\kappa_{\mathrm b,0}^{\mathrm{OBC}}
&=c_{\mathrm b}(N)L+o(L),\nonumber\\
c_{\mathrm b}(N)&=-\min_{x\in\mathbb R}\mathcal F_N(x)>0,
\label{eq:bond_exponential}
\end{align}
where $\mathcal F_N(x)=\frac{2}{\pi}\int_0^{\pi/2}d\varphi\,\ln|x-(2\cos\varphi)^{2/N}e^{\ii\pi/N}|^2$. For $N=3$, $c_{\mathrm b}(3)\simeq 0.1352026$: the exact response rises from $2.42$ at $L=20$ to $91.04$ at $L=50$ [matching our MPS data in Fig.~\ref{fig:fig5}(b)] and $5.53\times 10^4$ at $L=100$. An infinitesimal perturbation on a single physical bond thus produces an exponentially amplified first-order shift in the OBC ground-state energy. Because $B$ is unitary ($\eta_{\mathrm b,0}\le 1$), $\kappa_0^{\mathrm{OBC}}\ge\kappa_{\mathrm b,0}^{\mathrm{OBC}}$ also provides an exact exponential lower bound, $\liminf_{L\to\infty}L^{-1}\ln\kappa_0^{\mathrm{OBC}}\ge c_{\mathrm b}(N)>0$ for all $N>2$, while the stronger finite-size growth of $\kappa_0^{\mathrm{OBC}}$ is reflected in a decaying bond-alignment factor $\eta_{\mathrm b,0}$.

Under homogeneous PBC, all $L$ interaction bonds are equivalent. For any simple eigenstate, translation symmetry yields the exact identity
\begin{equation}
\kappa_{\mathrm b,n}^{\mathrm{PBC}}
=\frac{1}{L}
\left|\frac{\partial E_n^{\mathrm{PBC}}(\lambda)}
{\partial\lambda}\right|,
\label{eq:pbc_translation}
\end{equation}
where the derivative varies all $L$ bonds simultaneously. At the self-dual point $\lambda=1$ in $q=0$, periodic self-duality further equates the biorthogonal expectations of the field and interaction sums, yielding the exact finite-size identity (Appendix~\ref{app:bond_pbc})
\begin{equation}
\kappa_{\mathrm b,0}^{\mathrm{PBC}}=\frac{|E_0|}{2L}\le 1,
\label{eq:pbc_bond_exact}
\end{equation}
where the upper bound follows directly from $\|H\|_2\le 2L$. For $N=3$, our PBC MPS data decrease smoothly from $0.5840$ at $L=10$ to $0.5773$ at $L=50$---consistent with the thermodynamic energy-density extrapolation $|e_\infty|/2\simeq 0.5772$ of Ref.~\cite{AlcarazBatchelor2018}---while satisfying Eq.~\eqref{eq:pbc_bond_exact} to within $5\times 10^{-5}$ [Fig.~\ref{fig:fig5}(b)]. Hence, even though $\kappa_0^{\mathrm{PBC}}$ grows exponentially with $L$, translation symmetry and self-duality force $\eta_{\mathrm b,0}^{\mathrm{PBC}}$ to decay accordingly so that the single-bond response remains $O(1)$.

Table~\ref{tab:scaling} summarizes the scaling behaviors of the ground-state condition number $\kappa_0$, the largest simple-level condition number $\kappa_{\max}$, and the closing-bond response $\kappa_{\mathrm b,0}$ across both the fixed-$L$ (large-$N$) and fixed-$N$ (large-$L$) limits.

\begin{table}[t]
\caption{Summary of scaling behaviors for the ground-state condition number $\kappa_0$, the largest simple-level condition number $\kappa_{\max}$, and the physical closing-bond response $\kappa_{\mathrm b,0}$ at $\lambda=1$. Ground-state quantities refer to the neutral sector $q=0$, while $\kappa_{\max}$ denotes the strongest scaling among simple eigenvalues (excluding EPs).}
\label{tab:scaling}
\renewcommand{\arraystretch}{1.18}
\begin{ruledtabular}
\begin{tabular}{lccc}
Regime & $\kappa_0$ & $\kappa_{\max}$ & $\kappa_{\mathrm b,0}$ \\
\colrule
\multicolumn{4}{l}{\textit{Fixed $L$, increasing $N$}} \\
OBC & $\sim e^{\mathcal{O}(N)}$ & $\sim e^{\mathcal{O}(N)}$ & $\sim N^{2L-2}$ \\
PBC & $\mathcal{O}(1)$ & $\sim e^{\mathcal{O}(N)}$ & $\mathcal{O}(1)$ \\
\colrule
\multicolumn{4}{l}{\textit{Fixed $N>2$, increasing $L$}} \\
OBC & $\sim e^{\mathcal{O}(L^2)}$ & $\sim e^{\mathcal{O}(L^2)}$ & $\sim e^{\mathcal{O}(L)}$ \\
PBC & $\sim e^{\mathcal{O}(L)}$ & $\sim e^{\mathcal{O}(L^2)}$ & $\mathcal{O}(1)$
\end{tabular}
\end{ruledtabular}
\end{table}

\section{Exceptional points during boundary closure}
\label{sec:ep}

Beyond the infinitesimal response captured by $\kappa_{\mathrm b,n}$, tracking the full boundary interpolation $H(\mu)=H_{\mathrm{OBC}}-\mu B$ from $\mu=0$ to $\mu=1$ at $\lambda=1$ uncovers genuine spectral singularities. In contrast to previous realizations of EPs in the Baxter--Fendley chain using complex fields~\cite{EPBaxterModel}, all couplings remain real here: tuning the single boundary bond $\mu$ drives the finite chain through isolated EPs in $(0,1)$ before reaching symmetry-protected and composite exceptional degeneracies at the periodic endpoint ($\mu=1$).

\subsection{Boundary-induced level collisions for $0<\mu<1$}
\label{subsec:ep_intermediate}

In the neutral sector $q=0$ (and $q=N/2$ for even $N$), real couplings preserve $\mathcal R$ pseudo-Hermiticity, $H_q^\dagger(\mu)=\mathcal R H_q(\mu)\mathcal R$, together with the reflection--conjugation symmetry~\eqref{eq:PT_pairing}. Consequently, eigenvalues are either real or form complex-conjugate pairs. For any real simple level, the normalized right eigenvector carries a real inversion overlap
\begin{equation}
s_n=\langle\psi_n^{\mathrm R}|\mathcal R|\psi_n^{\mathrm R}\rangle,
\qquad
\kappa_n=\frac{1}{|s_n|}.
\label{eq:ep_inversion_signature}
\end{equation}
Because $s_n$ cannot vanish while the level remains real and simple, its sign is a conserved signature. Two real levels can merge and split into a complex-conjugate pair only if they carry opposite inversion signs~\cite{Melkani2023,Starkov2023}. Projecting $H_q(\mu)$ onto such a pair with metric $\mathrm{diag}(1,-1)$ gives the effective two-level model (Appendix~\ref{app:ep_signatures})
\begin{equation}
\begin{aligned}
H_{\mathrm{eff}} &= \begin{pmatrix}
e_1 & g\\
-g^* & e_2
\end{pmatrix},\\[4pt]
E_\pm &= \frac{e_1+e_2}{2}\pm\sqrt{\left(\frac{e_1-e_2}{2}\right)^2-|g|^2},
\end{aligned}
\label{eq:ep_two_mode_collision}
\end{equation}
with real $e_1,e_2$. Because the discriminant is strictly real, reaching a second-order exceptional point (EP2) at $|g|=|e_1-e_2|/2$ has codimension one---requiring only one real parameter $\mu$---where $s_\pm\to 0$ and $\kappa_\pm\to\infty$.

Because the number of real eigenvalues can change only when a complex-conjugate pair collides on the real axis, any difference in the real-eigenvalue count between OBC and PBC---unless absorbed by endpoint degeneracies right at $\mu=1$---requires crossing discrete EP2s within $\mu\in(0,1)$ [as seen in Fig.~\ref{fig:fig2}(a), where the complex OBC loop collapses onto distinct real levels under PBC]. For $N=3$, $L=6$, and $q=0$, we focus on two representative real-energy collisions in $\mu\in(0,1)$:
\begin{align}
\mu_{c,1}&=0.6319087270, & E_{\mathrm{EP},1}&=-4.5579313563,
\label{eq:mu_EP_outer}\\
\mu_{c,2}&=0.7814563640, & E_{\mathrm{EP},2}&=-2.3091334615.
\label{eq:mu_EP_inner}
\end{align}
At both critical couplings, a complex-conjugate pair merges on the real axis and splits into two real modes with opposite signs of $s_n$, accompanied by eigenvector coalescence with generalized nullities $(\nu_1,\nu_2)=(1,2)$ [Fig.~\ref{fig:fig6}(a); see Appendix~\ref{app:ep_jordan} for verification].

\begin{figure*}[t]
\centering
\plotfile[width=0.90\textwidth]{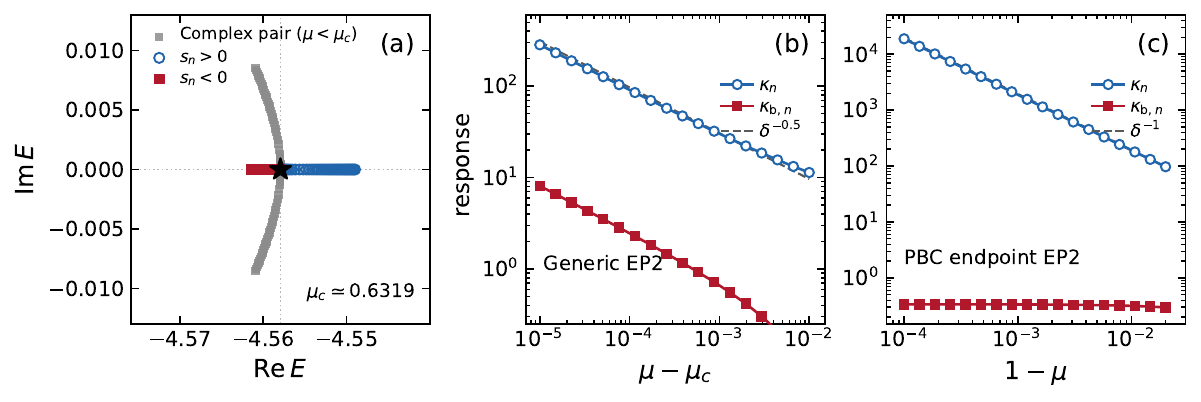}
\caption{Exceptional points during boundary closure for $N=3$, $L=6$, $q=0$, and $\lambda=1$. (a) Complex-plane trajectory ($\mathrm{Im}\,E$ vs.\ $\mathrm{Re}\,E$) of the colliding pair near the intermediate EP2 at $\mu_{c,1}\simeq 0.6319$ (black star): a complex-conjugate pair for $\mu<\mu_{c,1}$ (gray squares) merges and splits along the real axis into two modes with opposite inversion overlaps $s_n>0$ (blue open circles) and $s_n<0$ (red filled squares). (b) Scaling of the condition number $\kappa_n$ (blue open circles) and closing-bond response $\kappa_{\mathrm b,n}$ (red filled squares) versus $\delta=\mu-\mu_{c,1}$ near the generic intermediate EP2; both diverge as $\delta^{-1/2}$ (dashed line). (c) Same quantities approaching the duality-protected PBC endpoint EP2 at $E_{\mathrm{EP},-}=-3(2-\sqrt{2})^{1/3}$ versus $\delta=1-\mu$: $\kappa_n$ diverges as $\delta^{-1}$ (dashed line), whereas $\kappa_{\mathrm b,n}$ remains finite.}
\label{fig:fig6}
\end{figure*}

\subsection{Generic intermediate EP2 vs.\ duality-protected PBC endpoint EP2}
\label{subsec:ep_contrast}

The physical closing bond $B=Z_L^\dagger Z_1$ probes intermediate EP2s and PBC endpoint EP2s in qualitatively different ways. Near an isolated EP2 at $(\mu_c,E_c)$ with generalized eigenprojector $P_c$ and rank-one nilpotent part $D_c=(H(\mu_c)-E_c\id)P_c$ ($D_c^2=0$), varying $\delta\mu=\mu-\mu_c$ yields (Appendix~\ref{app:ep_splitting})
\begin{equation}
[E_+(\mu)-E_-(\mu)]^2=-4\,\delta\mu\,\operatorname{tr}(D_c B)+O(\delta\mu^2),
\label{eq:ep_main_discriminant}
\end{equation}
with $\kappa_\pm|E_+-E_-|\to\|D_c\|_2$.

At a generic intermediate EP2 ($\mu_{c,1}$ or $\mu_{c,2}$), $\operatorname{tr}(D_c B)\neq 0$ produces the standard square-root branch point,
\begin{equation}
|E_+-E_-|\propto|\mu-\mu_c|^{1/2},
\qquad
\kappa_\pm\propto\kappa_{\mathrm b,\pm}\propto|\mu-\mu_c|^{-1/2},
\label{eq:ep_generic_scaling}
\end{equation}
so both $\kappa_\pm$ and $\kappa_{\mathrm b,\pm}$ diverge while the bond-alignment factor $\eta_{\mathrm b,\pm}=\kappa_{\mathrm b,\pm}/\kappa_\pm$ approaches a finite constant [Fig.~\ref{fig:fig6}(b)].

At the homogeneous PBC endpoint ($\lambda=\mu=1$), however, $N=3$, $L=6$, $q=0$ hosts six isolated nonzero EP2s solving $E^6+108E^3+1458=0$, forming two threefold orbits with negative-real representatives
\begin{equation}
E_{\mathrm{EP},\pm}=-3\left(2\pm\sqrt{2}\right)^{1/3}
\label{eq:N3L6EPenergy}
\end{equation}
($E_{\mathrm{EP},+}\simeq -4.517361$, $E_{\mathrm{EP},-}\simeq -2.510158$). Here, periodic self-duality exchanges $A=\sum_j X_j$ and $C=\sum_j Z_j^\dagger Z_{j+1}$ while leaving $H_c=-A-C$ invariant. Together with $\operatorname{tr}(D_c H_c)=0$ and translation invariance across all $L$ bonds, this enforces (Appendix~\ref{app:ep_endpoint})
\begin{equation}
\operatorname{tr}(D_c B)=\frac{1}{L}\operatorname{tr}(D_c C)=0,
\label{eq:ep_endpoint_vanishing}
\end{equation}
canceling the square-root term in Eq.~\eqref{eq:ep_main_discriminant}. Approaching $E_{\mathrm{EP},-}$ from $\mu<1$ therefore gives linear splitting (with a nonvanishing second-order discriminant),
\begin{equation}
|E_+-E_-|\propto |1-\mu|,
\qquad
\kappa_\pm\propto |1-\mu|^{-1}.
\label{eq:ep_endpoint_scaling}
\end{equation}
Remarkably, although $\kappa_\pm$ diverges faster than at a generic EP2, the physical boundary response $\kappa_{\mathrm b,\pm}=|\partial_\mu E_\pm|$ remains finite because $\eta_{\mathrm b,\pm}\propto |1-\mu|$ [Fig.~\ref{fig:fig6}(c)]. This is the singular counterpart of Sec.~\ref{subsec:manybody_boundary}: a divergent intrinsic condition number need not yield a divergent response along a symmetry-protected perturbation direction.

\subsection{Commensurability and composite zero-energy EPs at PBC}
\label{subsec:ep_composite}

Because homogeneous PBC hosts many ordinary (diagonalizable) symmetry degeneracies and strongly non-normal states with large finite $\kappa_n$, defectiveness must be diagnosed within resolved symmetry sectors via the generalized nullities
\begin{equation}
\nu_r(E)=\dim\ker(H-E\id)^r=\sum_a\min(r,r_a),
\label{eq:ep_nullity_main}
\end{equation}
where $H|_E\simeq\bigoplus_a J_{r_a}(E)$ is the Jordan form at energy $E$. An ordinary degeneracy saturates at $r=1$, whereas $\nu_{r+1}>\nu_r$ proves a Jordan chain of length at least $r+1$.

Higher-order PBC EPs follow a clear commensurability rule. Whenever $d=\gcd(N,L)>1$ (with $s=N/d$), the unitary operator $\mathcal C_s=\prod_{j=1}^L Z_j^s X_j^{-js}$ generates a $d$-fold spectral rotation (Appendix~\ref{app:ep_jordan}),
\begin{equation}
\mathcal C_s H \mathcal C_s^{-1}=\exp\!\left(\frac{2\pi\ii}{d}\right)H,
\label{eq:spectral_rotation_main}
\end{equation}
for arbitrary $\lambda$ and $\mu$. Nonzero levels form $d$-fold orbits, while $E=0$ is the unique fixed point where spectral branches meet and form nilpotent Jordan chains. By contrast, coprime $(N,L)$ ($d=1$) lack this rotational protection for zero-energy Jordan chains.

\begin{table}[t]
\caption{Verified zero-energy Jordan decompositions $H|_{E=0}$ summed across all charge and momentum sectors (after resolving symmetry blocks) for representative commensurate periodic chains at $\lambda=\mu=1$. Here $m_{\mathrm{alg}}$ and $m_{\mathrm{geo}}=\nu_1(0)$ are the total algebraic and geometric multiplicities, and $J_r\equiv J_r(0)$ is a Jordan block of length $r$.}
\label{tab:zero_energy_jordan}
\renewcommand{\arraystretch}{1.2}
\begin{ruledtabular}
\begin{tabular}{ccc}
$(N,L)$ & $(m_{\mathrm{alg}},m_{\mathrm{geo}})$ & Jordan form at $E=0$ \\
\colrule
$(3,3)$ & $(6,5)$ & $J_2\oplus J_1^{\oplus 4}$ \\
$(3,6)$ & $(30,24)$ & $J_3\oplus J_2^{\oplus 4}\oplus J_1^{\oplus 19}$ \\
$(3,9)$ & $(144,106)$ & $J_4\oplus J_3^{\oplus 6}\oplus J_2^{\oplus 23}\oplus J_1^{\oplus 76}$ \\
$(4,6)$ & $(108,76)$ & $J_5^{\oplus 2}\oplus J_3^{\oplus 12}\oplus J_1^{\oplus 62}$ \\
$(6,3)$ & $(24,11)$ & $J_7\oplus J_5\oplus J_3\oplus J_2\oplus J_1^{\oplus 7}$
\end{tabular}
\end{ruledtabular}
\end{table}

Table~\ref{tab:zero_energy_jordan} lists verified zero-energy Jordan decompositions summed across all charge and momentum sectors at $\lambda=\mu=1$. Rather than a single high-order EP, $E=0$ hosts a composite exceptional degeneracy where Jordan chains of different lengths coexist with diagonalizable modes---for instance, $(N,L)=(3,6)$ has $(m_{\mathrm{alg}},m_{\mathrm{geo}})=(30,24)$ with maximal Jordan order three ($J_3\oplus J_2^{\oplus 4}\oplus J_1^{\oplus 19}$). Along $L=3,6,9$ at $N=3$, the maximal block size grows as $2,3,4$, and reaches $7$ at $(N,L)=(6,3)$ (occurring in the $q=3$ sector). While Eq.~\eqref{eq:spectral_rotation_main} pins $E=0$ as the rotational fixed point for any couplings, the specific zero-energy Jordan block sizes can still depend on $\lambda$ and $\mu$ (Appendix~\ref{app:ep_jordan}). Overall, boundary-driven and PBC EPs represent singular limits of the eigenvector nonorthogonality studied in Secs.~\ref{sec:skin} and~\ref{sec:manybody}, whereas strong boundaryless NHSE does not require fine-tuning to an EP.

\section{Conclusion and discussion}
\label{sec:conclusion}

In this work, we have explained the anomalous boundary sensitivity of the non-Hermitian Baxter--Fendley $Z_N$ spin chain through a boundaryless NHSE in many-body configuration space and its interplay with integrability. Within each fixed $Z_N$ charge sector, transforming to the local $X$ eigenbasis maps an $L$-site chain onto a directed hopping model with a nonuniform complex on-site potential on a periodic $(L-1)$-dimensional torus of linear size $N$. Under both physical OBC and PBC, this lattice has no geometric boundary; closing the spin chain simply adds an extra directed hopping channel at every site of the same torus. Whenever the cross commutator $[V_q,T^\dagger]+[T,V_q^\dagger]$ is nonzero, combining directed hopping with the nonuniform potential breaks normality and generates spatial growth and decay of eigenstate amplitudes, driving macroscopic eigenstate accumulation and left--right separation on a closed manifold. For $L=2$ and $L=3$, this boundaryless NHSE emerges under both OBC and PBC, with a unitary Weyl transformation exposing position-dependent asymmetric hopping on the PBC ring and the OBC conserved charge $J_2$ fixing two-channel interference on the $L=3$ torus.

Across many-body scales, commuting conserved quantities directly govern and constrain this nonorthogonality. Under OBC, the generating operator $\mathcal C(t)$ yields an exact ground-state spectral projector $\Pi_0$ built from bond-dimension-two MPO factors whose norm equals $\kappa_0^{\mathrm{OBC}}$, while under PBC the periodic generating operator $\mathcal C_{\mathrm{PBC}}(t)$ splits into two bond-dimension-two open-chain MPO branches that act as a Trotter-error-free polynomial MPS filter up to $L=50$. The OBC projector construction also establishes rigorous analytical upper bounds---$\ln\kappa_{\max}^{\mathrm{OBC}}=O(N)$ at fixed $L$ and $\ln\kappa_0^{\mathrm{OBC}}=O(L^2)$ at fixed $N$---with our $N=3$ MPS results showing $\ln\kappa_0^{\mathrm{OBC}}\propto L^2$ in sharp contrast to $\ln\kappa_0^{\mathrm{PBC}}\propto L$. At the same time, the exact OBC free-parafermion solution shows that the linear closing-bond response $\kappa_{\mathrm b,0}^{\mathrm{OBC}}$ grows exponentially with $L$ for all $N>2$, directly linking the boundaryless NHSE to anomalous boundary sensitivity and proving an exponential lower bound on $\kappa_0^{\mathrm{OBC}}$. By contrast, translation symmetry and self-duality at the PBC self-dual point strictly bound the single-bond response to $\kappa_{\mathrm b,0}^{\mathrm{PBC}}=|E_0|/(2L)\le 1$ even as $\kappa_0^{\mathrm{PBC}}$ grows exponentially. This distinction between intrinsic nonorthogonality and physical boundary response persists at boundary-induced EPs: while generic intermediate EP2s exhibit square-root splitting with divergent $\kappa_{\mathrm b,\pm}$, self-duality at the PBC endpoint EP2 cancels the leading square-root term, leaving $\kappa_{\mathrm b,\pm}$ finite despite a stronger $\delta^{-1}$ divergence of $\kappa_\pm$.

Several open questions merit further investigation. High-dimensional non-Hermitian lattices without translational invariance are generally not solvable, and severe NHSE-induced spectral instability often renders direct numerical diagonalization unreliable. Our analysis of the Baxter--Fendley $Z_N$ chain suggests that commuting conserved charges---acting either as exact spectral projectors or as Trotter-error-free polynomial MPO filters---provide a systematic route to designing and probing high-dimensional NHSE beyond standard numerical methods. Furthermore, it would be interesting to explore whether the integrable and free-parafermion structure of the Baxter--Fendley $Z_N$ model can be extended to open quantum many-body systems, such as Lindbladian dynamics, to obtain exact insight into dissipative spectra, steady states, and non-equilibrium relaxation.

\begin{acknowledgments}
X.F. is supported by Grant No.~12375027 from the National Natural Science Foundation of China.  S.C. was supported by the National Key Research and Development Program of China (Grant No.~2023YFA1406704) and the NSFC under Grants No.~12474287 and No.~T2121001. Numerical simulations were performed on the High-performance Computing Platform of Peking University.
\end{acknowledgments}

\appendix

\section{Directed ring for $L=2$}
\label{app:ring}

\subsection{OBC ground state and envelope}
\label{app:ring_obc}

For $L=2$, $q=0$, $\lambda=1$, and $\mu=0$, the right wavefunction obeys the exact one-step recurrence
\begin{equation}
\psi^{\mathrm R}(k+1)=g_k(E)\psi^{\mathrm R}(k),\qquad
g_k(E)=-E-2\cos\!\left(\frac{2\pi k}{N}\right).
\label{eq:app_L2_recursion}
\end{equation}
Periodicity around the ring requires $\prod_{k=0}^{N-1}g_k(E)=1$. Using the Chebyshev identity
\begin{equation}
\prod_{k=0}^{N-1}\left(x-2\cos\frac{2\pi k}{N}\right)
=2[T_N(x/2)-1],
\label{eq:app_L2_Chebyshev}
\end{equation}
we obtain $T_N(-E/2)=3/2$. The ground-state energy is therefore
\begin{equation}
E_0=-2\cosh\!\left(\frac{a}{N}\right),\qquad a=\operatorname{arcosh}\!\left(\frac{3}{2}\right).
\label{eq:app_L2_energy}
\end{equation}
Since all $g_k(E_0)$ are strictly positive, the ground-state amplitude grows or decays along the ring without phase cancellation.

For large $N$, set $x=k/N$ and write $\ln\psi_0^{\mathrm R}(k)=Nf(x)+o(N)$. Equation~\eqref{eq:app_L2_recursion} yields
\begin{equation}
f'(x)=\ln[4\sin^2(\pi x)],\qquad f(0)=0.
\label{eq:app_L2_envelope}
\end{equation}
The right amplitude grows across $1/6<x<5/6$, peaking at $x_{\mathrm R}=5/6$. By configuration inversion~\eqref{eq:configuration_inversion}, the left ground state peaks at $x_{\mathrm L}=1/6$. Because $f(1-x)=-f(x)$, the leading exponential factors cancel in the unnormalized overlap $\sum_k \psi_0^{\mathrm R}(N-k)\psi_0^{\mathrm R}(k)$, whereas the norm squared is dominated by $e^{2Nf(5/6)}$. Thus,
\begin{align}
\ln\kappa_0^{\mathrm{OBC}}
&=c_{\mathrm O}N+O(\ln N),\nonumber\\
c_{\mathrm O}&=2f(5/6)
=\int_{1/6}^{5/6}\!dx\,\ln[4\sin^2(\pi x)]
\simeq 0.646132.
\label{eq:app_L2_kappa_exponent}
\end{align}
The local curvature of $f(x)$ sets a Gaussian packet width of $O(\sqrt N)$, much smaller than the $O(N)$ separation between $k_{\mathrm L}$ and $k_{\mathrm R}$.

\subsection{PBC Weyl basis and hopping asymmetry}
\label{app:ring_pbc}

In the neutral sector $q=0$, homogeneous PBC ($\lambda=\mu=1$) is Hermitian with reciprocal hopping in the real potential $-2\cos(2\pi k/N)$, so $\kappa_0^{\mathrm{PBC}}=1$ for all $N$. Contrasting behavior arises in nontrivial charge sectors.

For even $N$ at $q=N/2$ and $\lambda=\mu=1$, the $X$-basis Hamiltonian is
\begin{align}
H_{N/2}^{\mathrm{PBC}}={}&\sum_{k=0}^{N-1}\biggl[2\ii\sin\!\left(\frac{2\pi k}{N}\right)\ket{k}_X{}_X\bra{k}\nonumber\\
&-\ket{k-1}_X{}_X\bra{k}-\ket{k+1}_X{}_X\bra{k}\biggr].
\label{eq:app_L2_halfcharge_X}
\end{align}
Introducing the unitary Weyl basis~\eqref{eq:weyl_basis},
\begin{equation}
\ket{m}_W=\frac{1}{\sqrt N}\sum_{k=0}^{N-1}
\exp\!\left(\frac{2\pi\ii mk}{N}-\frac{\ii\pi k^2}{N}\right)\ket{k}_X,
\label{eq:app_L2_U}
\end{equation}
the quadratic phase cancels from inner products, giving ${}_W\braket{m}{n}_W=\delta_{mn}$. Transforming Eq.~\eqref{eq:app_L2_halfcharge_X} into this basis eliminates the imaginary on-site potential and yields a pure nearest-neighbor hopping ring with position-dependent asymmetric amplitudes:
\begin{align}
\widetilde H&=\sum_{m=0}^{N-1}\bigl(a_m\ket{m+1}_W{}_W\bra{m}
+b_m\ket{m}_W{}_W\bra{m+1}\bigr),\nonumber\\
a_m&=1-e^{-\ii\theta_m},\qquad
b_m=-1-e^{\ii\theta_m},
\label{eq:app_L2_weyl_exact}
\end{align}
where $\theta_m=2\pi(m+1/2)/N$. The local hopping asymmetry obeys
\begin{equation}
\left|\frac{a_m}{b_m}\right|
=\left|\tan\frac{\pi(m+1/2)}{N}\right|,
\label{eq:app_L2_nonreciprocity}
\end{equation}
whose preferred direction reverses near $m=N/4$ and $3N/4$.

For even $N$, $\prod_{m=0}^{N-1}(a_m/b_m)=1$, so a periodic diagonal similarity transformation $S=\operatorname{diag}(s_m)$ with $|s_{m+1}/s_m|=\sqrt{|a_m/b_m|}$ symmetrizes the hopping magnitudes. Along $N=4M+2$,
\begin{align}
\ln\left|\frac{s_{m_+}}{s_{m_-}}\right|
&=\frac{1}{2}\sum_{m=m_-}^{m_+-1}
\ln\left|\tan\frac{\pi(m+1/2)}{N}\right|
=\gamma_{\mathrm P}N+O(\ln N),\nonumber\\
\gamma_{\mathrm P}&=\frac{1}{2}\int_{1/4}^{3/4}
\ln|\tan(\pi x)|\,dx
=\frac{G}{\pi}\simeq 0.291561,
\label{eq:app_L2_Catalan_gain}
\end{align}
where $m_-=\lfloor N/4\rfloor$, $m_+=\lfloor 3N/4\rfloor$, and $G$ is Catalan's constant. For strongly nonorthogonal real near-zero modes, global matching gives $\ln\kappa_{\mathrm{nz}}^{\mathrm{PBC}}\simeq \gamma_{\mathrm P}N$. Moreover, configuration inversion satisfies ${}_W\!\bra{m}\mathcal R\ket{n}_W=\delta_{m,-n\ (\mathrm{mod}\,N)}$, so $P_n^{\mathrm L}(m)=P_n^{\mathrm R}(-m)$ for simple real eigenvalues.

\section{Directed torus for $L=3$}
\label{app:torus}

\subsection{Joint eigenvalue equations from integrability}
\label{app:torus_joint}

For $L=3$ and OBC at $\lambda=1$, define
\begin{align}
h_1&=X_1,\quad h_2=Z_1^\dagger Z_2,\quad h_3=X_2,\nonumber\\
h_4&=Z_2^\dagger Z_3,\quad h_5=X_3,
\label{eq:app_L3_h_terms}
\end{align}
so that $H_{\mathrm{OBC}}=-\sum_{j=1}^{5}h_j$. Alongside $J_1=-H_{\mathrm{OBC}}$, the OBC chain possesses the commuting conserved charge~\cite{Fendley2014}
\begin{equation}
J_2=h_1h_3+h_1h_4+h_1h_5+h_2h_4+h_2h_5+h_3h_5,
\label{eq:app_L3_J2}
\end{equation}
with $[H_{\mathrm{OBC}},J_2]=0$. The ground state is a common eigenstate satisfying
\begin{align}
J_1\ket{\psi_0^{\mathrm R}}&=C_1\ket{\psi_0^{\mathrm R}},\qquad
J_2\ket{\psi_0^{\mathrm R}}=C_2\ket{\psi_0^{\mathrm R}},\nonumber\\
C_1&=\sum_{j=1}^3\epsilon_j,\qquad
C_2=\sum_{1\le i<j\le 3}\epsilon_i\epsilon_j,\nonumber\\
\epsilon_j&=\left[2\cos\frac{\pi j}{7}\right]^{2/N}.
\label{eq:app_L3_charge_values}
\end{align}

In the $X$-basis coordinates $(k_1,k_2)$ of the $q=0$ sector, set
\begin{equation}
a=\omega^{-k_1},\qquad b=\omega^{-k_2},\qquad
c=\omega^{k_1+k_2},\qquad abc=1,
\label{eq:app_L3_abc}
\end{equation}
and denote $\psi_{ij}=\psi_0^{\mathrm R}(k_1+i,k_2+j)$. The joint eigenvalue equations for $J_1$ and $J_2$ yield
\begin{align}
\psi_{1,-1}+\psi_{0,1}
&=(C_1-a-b-c)\psi_{0,0},\nonumber\\
\psi_{1,0}+c\psi_{1,-1}+a\psi_{0,1}
&=(C_2-ab-ac-bc)\psi_{0,0}.
\label{eq:app_L3_exact_joint}
\end{align}
The first relation fixes the sum of the two incoming hopping amplitudes, while the second, supplied by $J_2$, fixes their relative weights.

\subsection{Semiclassical packet and Hessian}
\label{app:torus_wkb}

At large $N$, introduce continuum coordinates $u=k_1/N$ and $v=k_2/N$ and the semiclassical ansatz
\begin{align}
\psi_0^{\mathrm R}(k_1,k_2)&\sim A(u,v)e^{NS(u,v)},\nonumber\\
p_1&=e^{\partial_u S-\partial_v S},\qquad
p_2=e^{\partial_v S}.
\label{eq:app_L3_wkb}
\end{align}
As $N\to\infty$, $C_1,C_2\to 3$, and Eq.~\eqref{eq:app_L3_exact_joint} reduces to
\begin{align}
p_1+p_2&=3-a-b-c,\nonumber\\
p_1p_2+cp_1+ap_2&=3-ab-ac-bc,
\label{eq:app_L3_wkb_constraints}
\end{align}
with $a=e^{-2\pi\ii u}$, $b=e^{-2\pi\ii v}$, and $c=e^{2\pi\ii(u+v)}$. Eliminating $p_2$ gives
\begin{equation}
p_1^2-(3-2a-b)p_1+\frac{(a-1)^3}{a}=0.
\label{eq:app_L3_quadratic}
\end{equation}

Along $k_2=0$ ($v=0$, $b=1$, $c=a^{-1}$), the two roots are $p_1=(1-a)(1\pm a^{-1/2})$. Selecting the branch matching the ground state gives
\begin{align}
p_1&=G(u)e^{\ii\theta(u)},\qquad
p_2=G(u)e^{-\ii\theta(u)},\nonumber\\
G(u)&=4\sin(\pi u)\cos\!\left(\frac{\pi u}{2}\right),\qquad
\theta(u)=\frac{\pi}{2}(1-u).
\label{eq:app_L3_gain_phase}
\end{align}
Since $e^{\partial_u S}=p_1p_2$, the step gain along $k_1$ is $|\psi_0^{\mathrm R}(k_1+1,0)/\psi_0^{\mathrm R}(k_1,0)|\simeq G^2(u)$. The larger root of $G(u)=1$ locates the right ground-state packet at
\begin{equation}
\bm k_{\mathrm R}\simeq N(u_{\mathrm R},0),\qquad
u_{\mathrm R}\simeq 0.761086.
\label{eq:app_L3_uR}
\end{equation}
Implicit differentiation of Eq.~\eqref{eq:app_L3_wkb_constraints} at $(u_{\mathrm R},0)$ gives the local Hessian
\begin{equation}
\left.\nabla^2\operatorname{Re}S\right|_{(u_{\mathrm R},0)}
\simeq\begin{pmatrix}
-14.7113&-7.35564\\
-7.35564&-9.65867
\end{pmatrix}.
\label{eq:app_L3_hessian_k}
\end{equation}
Both eigenvalues are negative and the off-diagonal entries are nonzero, confining the right ground state to a tilted two-dimensional Gaussian packet of width $O(\sqrt N)$.

\subsection{OBC condition-number scaling}
\label{app:torus_overlap}

Normalizing $\psi_0^{\mathrm R}(0,0)=1$ and writing $\ln|\psi_0^{\mathrm R}(uN,0)|=Nf(u)+o(N)$ along $k_2=0$, integrating the local gain $G^2(u)$ gives
\begin{equation}
f(u)=2\int_0^u\ln G(t)\,dt,\qquad
f(u_{\mathrm R})\simeq 0.917526.
\label{eq:app_L3_f_envelope}
\end{equation}
By configuration inversion,
\begin{equation}
\kappa_0^{\mathrm{OBC}}
=\frac{\sum_{\bm k}|\psi_0^{\mathrm R}(\bm k)|^2}
{\left|\sum_{\bm k}[\psi_0^{\mathrm R}(-\bm k)]^*\psi_0^{\mathrm R}(\bm k)\right|}.
\label{eq:app_L3_kappa_overlap}
\end{equation}
Global matching of the dominant semiclassical branch across the torus cancels the exponential factor in the unnormalized inversion overlap, yielding
\begin{equation}
\ln\kappa_0^{\mathrm{OBC}}
\simeq c_{\mathrm{sc}}N,\qquad
c_{\mathrm{sc}}=2f(u_{\mathrm R})\simeq 1.8351,
\label{eq:app_L3_conditional_c}
\end{equation}
in close agreement with exact diagonalization.

\subsection{Emergent Hermiticity of the PBC ground state}
\label{app:torus_pbc}

Under homogeneous PBC ($\mu=1$), the three hopping vectors satisfy $(-1,1)+(0,-1)+(1,0)=\bm 0$, so their first-order spatial derivatives cancel for smooth envelopes near $\bm k=\bm 0$. Expanding $d_0(\bm k)$ around the origin gives
\begin{equation}
d_0(\bm k)=3-\left(\frac{2\pi}{N}\right)^2
(k_1^2+k_1k_2+k_2^2)+O(|\bm k|^3/N^3).
\label{eq:app_L3_onsite_expansion}
\end{equation}
Combining this with the second-order hopping shifts yields the Hermitian effective Hamiltonian
\begin{align}
H_{\mathrm{eff}}^{\mathrm{PBC}}
\simeq{}&-6-\partial_{k_1}^2-\partial_{k_2}^2
+\partial_{k_1}\partial_{k_2}\nonumber\\
&+\left(\frac{2\pi}{N}\right)^2
(k_1^2+k_1k_2+k_2^2),
\label{eq:app_L3_oscillator}
\end{align}
whose ground state is the real Gaussian
\begin{equation}
\psi_0^{\mathrm{PBC}}(k_1,k_2)\propto
\exp\!\left[-\frac{2\pi}{\sqrt 3\,N}
(k_1^2+k_1k_2+k_2^2)\right].
\label{eq:app_L3_gaussian}
\end{equation}
Hence the left and right ground states coincide at leading order, giving $\kappa_0^{\mathrm{PBC}}\to 1$ and $\kappa_{\mathrm b,0}^{\mathrm{PBC}}\to 1$ as $N\to\infty$ at fixed $L=3$.

\section{Commuting-charge MPOs and MPS calculations}
\label{app:mps}

\subsection{Exact MPO representations for OBC and PBC}
\label{app:mpo}

For OBC at $\lambda=1$, define the $2L-1$ local terms $h_{2j-1}=X_j$ ($j=1,\ldots,L$) and $h_{2j}=Z_j^\dagger Z_{j+1}$ ($j=1,\ldots,L-1$). The commuting charges sum over products of mutually nonadjacent terms~\cite{Fendley2014}, so $\mathcal C(t)=\sum_{m=0}^L t^m J_m$ sums $t^{|I|}\prod_{a\in I}h_a$ over independent sets $I$ of the path graph $P_{2L-1}$, with eigenvalues~\eqref{eq:charge_eigenvalues}. Each factor $\mathcal C(t)$ admits the exact bond-dimension-two MPO representation~\eqref{eq:charge_MPO}, where the two virtual states track whether an interaction bond is open across the cut. Evaluating $\mathcal C(t)$ at $t_{js}=-\omega^s/\epsilon_j$ ($s=1,\ldots,N-1$) annihilates all excited occupation patterns while keeping $c_{\bm 0}(t_{js})\neq 0$, yielding the exact rank-one ground-state projector $\Pi_0$ in Eq.~\eqref{eq:ground_projector} with $\|\Pi_0\|_2=\kappa_0^{\mathrm{OBC}}$.

For homogeneous PBC at $\lambda=\mu=1$, adjoining the closing bond $h_{2L}=Z_L^\dagger Z_1$ arranges the $2L$ local generators on the cycle graph $C_{2L}$. Because nonadjacent generators on $C_{2L}$ commute pairwise, summing over all independent sets of $C_{2L}$ defines a periodic generating operator $\mathcal C_{\mathrm{PBC}}(t)=\sum_{m=0}^L t^m I_m$ ($I_0=\id$, $I_1=-H_{\mathrm{PBC}}$) satisfying $[\mathcal C_{\mathrm{PBC}}(t),\mathcal C_{\mathrm{PBC}}(t')]=0$ and $[\mathcal C_{\mathrm{PBC}}(t),H_{\mathrm{PBC}}]=0$ for all $t,t'$~\cite{LiLiuBatchelor2026}. Partitioning the independent sets of $C_{2L}$ according to whether $h_{2L}$ is excluded or included (which forbids its adjacent field terms $h_1=X_1$ and $h_{2L-1}=X_L$ while permitting the bulk bonds $h_2=Z_1^\dagger Z_2$ and $h_{2L-2}=Z_{L-1}^\dagger Z_L$, with $Z_1 Z_1^\dagger=\id$ and $Z_L Z_L^\dagger=\id$) splits $\mathcal C_{\mathrm{PBC}}(t)$ into two bond-dimension-two open-chain MPO branches:
\begin{align}
\mathcal C_{\mathrm{PBC}}(t)
={}&\mathcal C(t)
+\bigl(tZ_1\ \ t^2\id\bigr)\,W_2(t)\cdots W_{L-1}(t)
\begin{pmatrix}Z_L^\dagger\\\id\end{pmatrix},
\label{eq:app_pbc_commuting_mpo}
\end{align}
where $W_j(t)$ ($j=2,\ldots,L-1$) are the exact same $2\times 2$ bulk MPO tensors as in Eq.~\eqref{eq:charge_MPO}.

\subsection{Analytical upper bounds on OBC condition numbers}
\label{app:projector_bounds}

The projector construction extends to any simple OBC eigenstate with free-parafermion occupation pattern $\bm p=(p_1,\ldots,p_L)$:
\begin{equation}
\Pi_{\bm p}=\prod_{j=1}^L\prod_{\substack{0\le s<N\\s\neq p_j}}
\frac{\mathcal C(-\omega^{-s}/\epsilon_j)}
{c_{\bm p}(-\omega^{-s}/\epsilon_j)},
\qquad \|\Pi_{\bm p}\|_2=\kappa_{\bm p}^{\mathrm{OBC}}.
\label{eq:app_general_projector}
\end{equation}
At fixed $L$, all $\epsilon_j$ are uniformly bounded above and away from zero in $N$. Since $\|h_a\|_2=1$, $\|\mathcal C(t)\|_2\le (1+|t|)^{2L-1}$, so a constant $M_L$ independent of $N$ and $\bm p$ bounds every numerator norm. Using $\prod_{s=0}^{N-1}(1-r\omega^s)=1-r^N$, the denominator obeys
\begin{equation}
\prod_{s\neq p_j}|c_{\bm p}(-\omega^{-s}/\epsilon_j)|
=N\prod_{\ell\neq j}
\frac{|1-(\epsilon_\ell/\epsilon_j)^N|}
{|1-(\epsilon_\ell/\epsilon_j)\omega^{p_\ell-p_j}|}.
\label{eq:app_projector_denominator}
\end{equation}
At fixed $L$, $(\epsilon_\ell/\epsilon_j)^N$ is independent of $N$ and distinct from $1$ for $\ell\neq j$, while each factor in the denominator of Eq.~\eqref{eq:app_projector_denominator} is uniformly bounded above. Thus there exists $D_L>0$ independent of $N$ and $\bm p$ such that
\begin{equation}
\kappa_{\bm p}^{\mathrm{OBC}}=\|\Pi_{\bm p}\|_2\le \frac{M_L^{L(N-1)}}{D_L N^L},
\label{eq:app_fixed_l_bound}
\end{equation}
proving the uniform upper bound $\ln\kappa_{\max}^{\mathrm{OBC}}=O(N)$ at any fixed $L$.

At fixed $N>2$ and growing $L$, define $d_N=\min_{1\le s<N,\,r\ge 0}|1-r\omega^s|>0$, so that $|c_{\bm 0}(t_{js})|\ge d_N^L$. Using $\epsilon_j\le 2^{2/N}$ and $\prod_{j=1}^L\epsilon_j=1$ gives $\sum_{j=1}^L\ln(1+\epsilon_j^{-1})=\sum_{j=1}^L\ln(1+\epsilon_j)\le L\ln(1+2^{2/N})$. Bounding Eq.~\eqref{eq:ground_projector} therefore yields
\begin{equation}
\ln\kappa_0^{\mathrm{OBC}}\le (N-1)\bigl[L(2L-1)\ln(1+2^{2/N})-L^2\ln d_N\bigr],
\label{eq:app_fixed_n_bound}
\end{equation}
proving $\ln\kappa_0^{\mathrm{OBC}}=O(L^2)$ at fixed $N$.

\subsection{MPS evaluation and convergence checks}
\label{app:mps_checks}

Under OBC, starting from the neutral, translation-invariant product state $\ket{\Omega}=\ket{0}_X^{\otimes L}$, we apply the $L(N-1)$ MPO factors of $\Pi_0$ sequentially and compress the resulting MPS $\ket v=\Pi_0\ket\Omega\propto\ket{\psi_0^{\mathrm R}}$ to bond dimension $\chi$ after each step while tracking its overall normalization, exploiting $Z_N$ charge conservation on all virtual bonds. Because configuration inversion acts locally as $\mathcal R\ket{k_1,\ldots,k_L}_X=\ket{-k_1,\ldots,-k_L}_X$ and obeys $\Pi_0^2=\Pi_0$, $\Pi_0^\dagger=\mathcal R\Pi_0\mathcal R$, and $\mathcal R\ket\Omega=\ket\Omega$, we have $\langle v|\mathcal R|v\rangle=\langle\Omega|v\rangle$, yielding the two estimators in Eq.~\eqref{eq:MPS_kappa} as an internal consistency check on MPS truncation. Benchmarked against left--right ED at $N=3$, $L=12$, and $\lambda=1$ ($\kappa_0^{\mathrm{OBC}}=2.4673240667696$), the relative error drops from $4.4\times 10^{-3}$ at $\chi=8$ to $2.1\times 10^{-13}$ at $\chi=128$ (with eigenvector residual $\|(H-E_0)\psi_0^{\mathrm R}\|=2.9\times 10^{-11}$). Across $10\le L\le 50$ at $N=3$ (bond dimensions up to $\chi=256$), representative converged OBC values are $\kappa_0^{\mathrm{OBC}}\simeq 7.010$ ($L=20$), $44.72$ ($L=30$), $551.5$ ($L=40$), and $1.201\times 10^4$ ($L=50$), where the inversion and projector estimators agree to within $5\times 10^{-5}$ and the MPS closing-bond responses match the exact integral~\eqref{eq:bond_integral} to within $4\times 10^{-6}$.

Under PBC, we initialize $\ket\Omega=\ket{0}_X^{\otimes L}$ (which lies in the $q=0$, zero-momentum sector) and iterate the commuting-charge polynomial filter~\eqref{eq:app_pbc_commuting_mpo} by cycling positive real parameters $t\in\{0.85,0.55,0.35\}$, followed by a parameter-stability check with $t\in\{0.70,0.45\}$. At each step, the two bond-dimension-two MPO branches in Eq.~\eqref{eq:app_pbc_commuting_mpo} are applied and compressed within contiguous $Z_N$ charge blocks before summing and recompressing to bond dimension $\chi$, keeping peak intermediate bond dimensions at $2\chi$ without any Trotter step-size error. Increasing $\chi$ along a convergence ladder up to $\chi=256$ across $10\le L\le 50$ at $N=3$ reproduces ED at $L=10,12$ ($\kappa_0^{\mathrm{PBC}}\simeq 1.2748, 1.3653$) and yields converged long-chain values $\kappa_0^{\mathrm{PBC}}\simeq 1.7715$ ($L=20$), $2.3998$ ($L=30$), $3.2225$ ($L=40$), and $4.3172$ ($L=50$). Across all sizes up to $L=50$, the relative right-eigenvector residuals $\|(H-E_0)v\|/[(1+|E_0|)\|v\|]$ and self-duality deviations $|\kappa_{\mathrm b,0}^{\mathrm{PBC}}-|E_0|/(2L)|$ remain below $5\times 10^{-5}$, while $Z_N$ charge violations stay at machine precision.

\subsection{Transfer-matrix interpretation of the PBC overlap}
\label{app:pbc_transfer}

Assuming the PBC ground state is well described by a translationally invariant periodic MPS with fixed or effectively converged local tensors $A^k$ in the $X$ basis, its configuration-inverted partner $\mathcal R\ket\psi$ has tensors $A^{-k}$. Defining the ordinary and mixed transfer matrices
\begin{equation}
T=\sum_{k=0}^{N-1} \overline{A^k}\otimes A^k,
\qquad
T_{\mathcal R}=\sum_{k=0}^{N-1} \overline{A^{-k}}\otimes A^k,
\label{eq:app_pbc_transfer}
\end{equation}
the normalized inversion overlap is $\operatorname{tr}(T_{\mathcal R}^L)/\operatorname{tr}(T^L)$. Normalizing $T$ to spectral radius $1$, whenever $T_{\mathcal R}$ has a unique dominant eigenvalue of modulus $0<r<1$, we obtain $\ln\kappa_0^{\mathrm{PBC}}=-L\ln r+o(L)$, providing a natural mechanism for extensive logarithmic growth under PBC.

\section{Physical closing-bond response}
\label{app:bond}

\subsection{OBC: fixed $L$ and $N\to\infty$}
\label{app:bond_large_n}

In the exact OBC ground-state closing-bond response~\eqref{eq:bond_integral}~\cite{LiLiuBatchelor2026}, set $\beta_j=2\ln[2\cos(\pi j/(2L+1))]$ so that $\epsilon_j=e^{\beta_j/N}$. Rescaling $x=1+z/N$ at fixed $L$ and large $N$ gives
\begin{equation}
|x-\epsilon_j e^{\ii\pi/N}|^2
=N^{-2}\bigl[(z-\beta_j)^2+\pi^2\bigr]+O(N^{-3}).
\label{eq:app_bond_fixedL_expansion}
\end{equation}
Combining $N^{2L}$ from the denominator with $N^{-1}$ each from $dx$ and $\sin(\pi/N)/\pi$ yields
\begin{equation}
\kappa_{\mathrm b,0}^{\mathrm{OBC}}\sim C_L N^{2L-2},
\qquad
C_L=\int_{-\infty}^{\infty}\frac{dz}
{\displaystyle\prod_{j=1}^L\bigl[(z-\beta_j)^2+\pi^2\bigr]}.
\label{eq:app_bond_power}
\end{equation}
For $L=2$ ($\beta_{1,2}=\pm a$), this gives $C_2=1/[2(\pi^2+a^2)]\simeq 0.046314$; for $L=3$, numerical integration yields $C_3\simeq 0.00301411$.

\subsection{OBC: fixed $N>2$ and $L\to\infty$}
\label{app:bond_large_l}

Rewriting Eq.~\eqref{eq:bond_integral} as $\kappa_{\mathrm b,0}^{\mathrm{OBC}}=\frac{\sin(\pi/N)}{\pi}\int_{-\infty}^\infty e^{-L\mathcal F_{N,L}(x)}\,dx$ with $\mathcal F_{N,L}(x)=\frac{1}{L}\sum_{j=1}^L\ln|x-\epsilon_j e^{\ii\pi/N}|^2$, the $L\to\infty$ rate function is
\begin{equation}
\mathcal F_N(x)=\frac{2}{\pi}\int_0^{\pi/2}\!d\varphi\,
\ln\left|x-(2\cos\varphi)^{2/N}e^{\ii\pi/N}\right|^2.
\label{eq:app_bond_rate}
\end{equation}
Laplace's method gives $\lim_{L\to\infty}L^{-1}\ln\kappa_{\mathrm b,0}^{\mathrm{OBC}}=c_{\mathrm b}(N)=-\min_{x\in\mathbb R}\mathcal F_N(x)$. Since $\int_0^{\pi/2}\ln(2\cos\varphi)\,d\varphi=0$, we have $\mathcal F_N(0)=0$, whereas for all $N>2$,
\begin{equation}
\mathcal F_N'(0)=-\frac{4\cos(\pi/N)}{\pi}\int_0^{\pi/2}\frac{d\varphi}{(2\cos\varphi)^{2/N}}<0.
\label{eq:app_bond_derivative}
\end{equation}
Hence $\mathcal F_N(x)<0$ for small positive $x$, proving $c_{\mathrm b}(N)>0$ [with $c_{\mathrm b}(3)\simeq 0.1352026$]. Since $B$ is unitary, $\kappa_0^{\mathrm{OBC}}\ge\kappa_{\mathrm b,0}^{\mathrm{OBC}}$ establishes $\liminf_{L\to\infty}L^{-1}\ln\kappa_0^{\mathrm{OBC}}\ge c_{\mathrm b}(N)>0$.

\subsection{PBC: translation symmetry and self-duality}
\label{app:bond_pbc}

Write $H_{\mathrm{PBC}}(\lambda)=-A-\lambda C$, with $A=\sum_{j=1}^L X_j$, $C=\sum_{j=1}^L Y_j$, and $Y_j=Z_j^\dagger Z_{j+1}$ ($Y_L=B$). For any simple eigenstate, lattice translation symmetry equates the biorthogonal expectation values of all $L$ bonds:
\begin{equation}
\frac{\langle\psi_n^{\mathrm L}|Y_j|\psi_n^{\mathrm R}\rangle}
{\langle\psi_n^{\mathrm L}|\psi_n^{\mathrm R}\rangle}
=\frac{1}{L}\frac{\langle\psi_n^{\mathrm L}|C|\psi_n^{\mathrm R}\rangle}
{\langle\psi_n^{\mathrm L}|\psi_n^{\mathrm R}\rangle},
\label{eq:app_PBC_equal_bonds}
\end{equation}
which yields Eq.~\eqref{eq:pbc_translation}. At the self-dual point $\lambda=1$ in the neutral sector $q=0$, the constraint $\prod_{j=1}^L X_j=\prod_{j=1}^L Y_j=\id$ enables the unitary periodic duality $X_j\mapsto Y_j$, $Y_j\mapsto X_{j+1}$, which exchanges $A$ and $C$ while leaving $H_c=-A-C$ invariant. For a simple eigenstate, its spectral projector is duality invariant, enforcing $\langle A\rangle_{\mathrm{bi}}=\langle C\rangle_{\mathrm{bi}}=-E_n/2$. Combining this with Eq.~\eqref{eq:app_PBC_equal_bonds} gives the exact finite-size identity $\kappa_{\mathrm b,0}^{\mathrm{PBC}}=|E_0|/(2L)\le 1$, where the bound follows directly from $\|H_c\|_2\le 2L$.

\section{Exceptional points and perturbation directions}
\label{app:ep}

\subsection{Inversion signatures and real-level collisions}
\label{app:ep_signatures}

In the $q=0$ sector with real couplings, $H^\dagger\mathcal R=\mathcal R H$ ($\mathcal R^2=\id$) implies that right eigenstates obey
\begin{equation}
(E_m^*-E_n)\langle\psi_m^{\mathrm R}|\mathcal R|\psi_n^{\mathrm R}\rangle=0.
\label{eq:app_EP_Rorth}
\end{equation}
For a normalized real simple level, $s_n=\langle\psi_n^{\mathrm R}|\mathcal R|\psi_n^{\mathrm R}\rangle$ is real and nonzero with $\kappa_n=|s_n|^{-1}$, and its sign is conserved along any real simple branch. Two real levels can merge and split into a complex-conjugate pair only if they carry opposite signs of $s_n$~\cite{Melkani2023,Starkov2023}. Choosing a basis in this two-level subspace with inversion metric $\operatorname{diag}(1,-1)$ yields the projected pseudo-Hermitian Hamiltonian~\eqref{eq:ep_two_mode_collision}, where increasing $|g|$ to $|e_1-e_2|/2$ produces a codimension-one EP2.

\subsection{Perturbation-dependent splitting at an EP2}
\label{app:ep_splitting}

Near an isolated EP2 at $(\mu_c,E_c)$ with generalized eigenprojector $P_c$ and rank-one nilpotent part $D_c=(H(\mu_c)-E_c\id)P_c$ ($D_c\neq 0$, $D_c^2=0$), let $P(\mu)$ project onto the two-level spectral subspace. For two distinct nearby eigenvalues $E_\pm(\mu)$, the individual eigenprojectors satisfy
\begin{equation}
P_\pm(\mu)=\frac{[H(\mu)-E_\mp(\mu)\id]P(\mu)}{E_\pm(\mu)-E_\mp(\mu)}
=\pm\frac{D_c}{E_+(\mu)-E_-(\mu)}+O(1),
\label{eq:app_EP_projector}
\end{equation}
yielding $\lim_{\mu\to\mu_c}\kappa_\pm(\mu)|E_+(\mu)-E_-(\mu)|=\|D_c\|_2$. Under the closing-bond perturbation $\delta H=-\delta\mu\,B$ ($\delta\mu=\mu-\mu_c$), the leading term of the two-level discriminant is controlled by $\operatorname{tr}(D_c B)$, giving Eq.~\eqref{eq:ep_main_discriminant}. When $\operatorname{tr}(D_c B)\neq 0$, as at the generic intermediate EP2s in Eqs.~\eqref{eq:mu_EP_outer} and~\eqref{eq:mu_EP_inner}, $|E_+-E_-|\propto|\delta\mu|^{1/2}$ and $\kappa_\pm,\kappa_{\mathrm b,\pm}\propto|\delta\mu|^{-1/2}$.

\subsection{Duality-protected PBC endpoint}
\label{app:ep_endpoint}

At the self-dual PBC endpoint ($\lambda=\mu=1$, $q=0$), both $P_c$ and $D_c$ of an isolated EP2 are invariant under lattice translation and the periodic duality $X_j\mapsto Y_j$, $Y_j\mapsto X_{j+1}$ (Appendix~\ref{app:bond_pbc}). With $H_c=-A-C$, duality enforces $\operatorname{tr}(D_c A)=\operatorname{tr}(D_c C)$, while nilpotency ($D_c^2=0$) gives $\operatorname{tr}(D_c H_c)=0$. Combining this with translation invariance across all $L$ bonds yields Eq.~\eqref{eq:ep_endpoint_vanishing}, canceling the leading square-root term in Eq.~\eqref{eq:ep_main_discriminant}. Approaching the $N=3$, $L=6$ endpoint EP2 at $E_{\mathrm{EP},-}=-3(2-\sqrt 2)^{1/3}$ from $\mu<1$ therefore yields linear splitting $|E_+-E_-|\propto|1-\mu|$, $\kappa_\pm\propto|1-\mu|^{-1}$, and finite $\kappa_{\mathrm b,\pm}=O(1)$.

\subsection{Generalized nullities and spectral rotation}
\label{app:ep_jordan}

Within resolved symmetry sectors, defectiveness is diagnosed via the generalized nullities $\nu_r(E)=\dim\ker(H-E\id)^r=\sum_a\min(r,r_a)$, where $\nu_r-\nu_{r-1}$ counts Jordan blocks of length at least $r$ ($\nu_0=0$). For $N=3$, $L=6$, and $q=0$, the intermediate collisions at $\mu_{c,1}$ and $\mu_{c,2}$ have $(\nu_1,\nu_2)=(1,2)$, confirming isolated EP2s. At homogeneous PBC ($\lambda=\mu=1$), the zero-energy subspace has nullities $(24,29,30,30)$, corresponding to $J_3(0)\oplus J_2(0)^{\oplus 4}\oplus J_1(0)^{\oplus 19}$, while the six nonzero EP2s lie at $E_{\mathrm{EP},\pm,j}=-3(2\pm\sqrt 2)^{1/3}\omega^j$ ($j=0,1,2$).

Whenever $d=\gcd(N,L)>1$ with $s=N/d$, the unitary operator $\mathcal C_s=\prod_{j=1}^L Z_j^s X_j^{-js}$ commutes with $Q$ and multiplies every field term $X_j$, bulk bond $Z_j^\dagger Z_{j+1}$, and closing bond $Z_L^\dagger Z_1$ (since $Ls\equiv 0\pmod N$) by $\omega^s=e^{2\pi\ii/d}$, proving Eq.~\eqref{eq:spectral_rotation_main} for arbitrary $\lambda$ and $\mu$. Nonzero eigenvalues thus form $d$-fold rotational orbits, while $E=0$ is the unique fixed point where spectral branches meet and form nilpotent Jordan chains. While Eq.~\eqref{eq:spectral_rotation_main} holds for arbitrary couplings, the resulting zero-energy Jordan block sizes can depend on $\lambda$ and $\mu$ [for example, moving from $\lambda=\mu=1$ to $\lambda=\mu=2$ leaves the $(N,L)=(3,6)$, $q=0$ zero-energy nullities $(24,29,30,30)$ unchanged, whereas in $(N,L)=(6,3)$, $q=3$ the largest zero-energy block reduces from $J_7(0)$ to $J_2(0)$].

\bibliographystyle{apsrev4-2}
\bibliography{ref}

\end{document}